\documentclass[prb,twocolumn,showpacs]{revtex4}
\usepackage{graphicx,epsfig}
\renewcommand{\vec}[1]{{\mathbf #1}}
\newcommand{\calM}{{\mathcal M}}
\newcommand{\epstil}{\tilde{\epsilon}}
\newcommand{\omtil}{\tilde{\omega}}
\newcommand{\rhol}{\rho_{\mathrm{L}}}
\newcommand{\sech}{\mathrm{sech}}
\begin{document}

\title{BCS-BEC crossover in a system of microcavity polaritons}
\author{Jonathan Keeling}
\email{jmjk2@cam.ac.uk}
\author{P.~R.~Eastham}
\author{M.~H.~Szymanska}
\author{P.~B.~Littlewood}
\affiliation{Cavendish Laboratory, Madingley Road, Cambridge CB3 OHE, U.K.}

\date{\today}

\begin{abstract}
  We investigate the thermodynamics and signatures of a polariton
  condensate over a range of densities, using a model of microcavity
  polaritons with internal structure.
  We determine a phase diagram for this system including fluctuation
  corrections to the mean-field theory.
  At low densities the condensation temperature, $T_c$, behaves like
  that for point bosons.
  At higher densities, when $T_c$ approaches the Rabi splitting, $T_c$
  deviates from the form for point bosons, and instead approaches
  the result of a BCS-like mean-field theory.
  This crossover occurs at densities much less than the Mott density.
  We show that current experiments are in a density range where the
  phase boundary is described by the BCS-like mean-field boundary.
  We investigate the influence of inhomogeneous broadening and detuning
  of excitons on the phase diagram.
\end{abstract}
\maketitle

\section{Introduction}
\label{sec:introduction}

There have been many recent experiments with the aim of observing Bose
condensation of polaritons in two-dimensional microcavities.
Recent experimental progress includes nonlinear increase of the
occupation of the ground
state\cite{dang98:_stimul,yamamoto02:_condensation},
sub-thermal second order coherence\cite{yamamoto02:_condensation},
changes to the angular
dispersion of polariton luminescence\cite{deng03:_polar,weihs04:_polar,richard04:_angle_cdte_ii_vi}
increased population of low momentum polaritons\cite{deng03:_polar},
and stimulated processes in resonantly pumped
cavities\cite{baas05:_quant,savasta04:_quant,baumberg02:_polar,savvidis00:_angle,baumberg00:_param}.
Such experiments however do not provide unambiguous evidence for the
observation of a polariton condensate as distinct from, {\it e.g.}\  a
novel kind of laser.
Theoretical predictions of the detailed properties of polariton
condensation are therefore of considerable interest.
In this paper, we consider the form of the phase boundary, and
experimental signatures of condensation.
We include the internal structure of polaritons, and fermionic structure
of excitons.

The phase boundary at low densities may be described by a model of
weakly interacting
bosons\cite{kavokin03:_polar,laussy04:_spont,tassone97:_bottl}.
At higher densities, internal polariton structure becomes important,
and the phase boundary approaches the result of a BCS-like mean-field
theory.
For a zero-dimensional cavity, the mean-field theory is correct at all
densities, but in a two-dimensional cavity, fluctuations cause a
crossover to the weakly interacting boson limit at low densities.
This crossover is shown in the top panel of the phase diagram,
figure~\ref{fig:phasediag}.
The crossover scale is set by the wavelength of light, rather than
average exciton separation, and so occurs at densities much less than
the Mott density.
Figures~\ref{fig:phasediag} and~\ref{fig:mu-rho-T} also show the effect
of detuning the exciton below the photon on the phase boundary.
As in the mean-field case\cite{eastham01:_bose}, detuning 
can lead to a multi-valued phase boundary.
However, with fluctuations this multi-valued structure occurs for
smaller detunings than are required for the mean-field case.

We also discuss a number of experimentally testable signatures of a
coherent state of polaritons.
These include dramatic changes in the luminescence and absorption
spectra, as shown in figures~\ref{fig:simple_spectrum}
and~\ref{fig:dos}, and in the angular distribution of radiation
emitted from the condensate, as shown in figure~\ref{fig:nofk}.
These signatures are a consequence of a coherent photon field, and of
the presence of the Goldstone mode associated with symmetry breaking.
In a previous publication\cite{keeling04:_polar}, some of these
results were presented in the absence of exciton-photon detuning.
In this paper we present in full the calculation which led to those
results, and extend our results to include detuning and inhomogeneous
broadening of the excitons.

We consider a model of localised excitons coupled to a continuum of
radiation modes confined in a two-dimensional microcavity.
This provides an extension of previous
studies\cite{eastham00:_bose,eastham01:_bose} of the mean-field
(zero-dimensional) system.
A model of localised excitons is motivated by systems such as organic
semiconductors\cite{lidzey99:_room,lidzey98:_stron}, quantum
dots\cite{woggon03:_dotindot,reithmaler04:_stron,yoshi04:_vacuum} and
disordered quantum wells\cite{hess94}.
In addition, the predictions of such a model are expected to be
similar to those for a model of mobile excitons, since a typical
exciton mass is several orders of magnitude larger than the typical
photon mass.

We study the behaviour of this model in thermal equilibrium.
Although current experiments may remain far from equilibrium, there
are several reasons to study the equilibrium behaviour.
As the quality of microcavity fabrication, and particular the quality
of mirrors improve\cite{pawlis04:_prper,tawara04}, experiments can be
expected to reach states closer to thermal equilibrium.
Further, the nature of experimental signatures for coherence close to
equilibrium are expected to be similar to those in thermal
equilibrium.
Finally, the equilibrium distribution may be considered a limiting
case of the non-equilibrium problem, when pumping and decay rates are
taken to zero, so the equilibrium case is thus instructive in
approaching the non-equilibrium problem.

The Hamiltonian we study is similar to the Holland-Timmermans
Hamiltonian\cite{holland01:_reson_super_quant_degen_fermi_gas,timmermans99:_feshb_bose_einst}
studied in the context of Feshbach resonances in atomic gases.
However, there are a number of important differences between the two
models; most notably the absence of a direct four-fermion interaction,
and the bare fermion density of states.
How these differences affect the mean-field results is discussed in
section~\ref{sec:comp-mean-field}, and their effect on the fluctuation
spectrum in
sections~\ref{sec:effect-acti-fluct} and~\ref{sec:poles-greens-funct}.
Because our system is two-dimensional, it is necessary to study
fluctuations in the presence of a condensate.
Such fluctuations have recently been studied by Ohashi and
Griffin\cite{ohashi03:_super_fermi_feshb} in the context of the
Feshbach resonances.
Despite differences in the models, we disagree with their use of
partial derivatives of the free energy w.r.t.\ chemical potential,
rather than full derivatives.
As we show, their approach neglects terms of the same order as those
that are included.

The rest of this paper is organised as follows.
In section~\ref{sec:model} we introduce the Hamiltonian for our model,
and discuss how it will be treated.
Section~\ref{sec:greens-funct-fluct} reproduces the mean-field
results presented elsewhere, and then discusses the effective
action, and resulting spectrum of fluctuations about it.  
From the fluctuation spectrum, a number of experimental signatures are
also discussed.

In section~\ref{sec:fluct-corr-dens} we explain how to calculate
fluctuation corrections to the mean-field phase diagram.
Some of the issues in this section are associated with the specific
details of our model, or with fluctuations in two dimensions, but
others are relevant to the calculation of fluctuation corrections
in general.
Section~\ref{sec:new-phase-boundary} presents the results of
including fluctuations, and discusses the effects of detuning
and inhomogeneous broadening on the phase boundary.  
The discussion presented in section~\ref{sec:new-phase-boundary} does
not require the details presented in
section~\ref{sec:fluct-corr-dens}, and readers not interested in the
theoretical explication may move directly to
section~\ref{sec:new-phase-boundary}.
Finally, in section~\ref{sec:conclusions} we summarise our
conclusions.

The calculation of the thermal Green's function for fluctuations in
the condensed state includes terms proportional to $\delta_{\omega}$.
In the appendix, we discuss in general how such terms may arise in the
thermal Green's function, but vanish after analytic continuation and
so are not present in the retarded Green's function.

\section{The model}
\label{sec:model}

Our model describes localised two-level systems, coupled to a
continuum of radiation modes in a two-dimensional microcavity.
Considering only two levels describes a hard-core repulsion between
excitons on a single site.
We do not consider a static Coulomb interaction between different
sites.
The two-level systems may either be represented as fermions with an
occupancy constraint, as will be described below, or as spins with
magnitude $|\vec{S}|=1/2$.
In the latter case, the generalised Dicke
Hamiltonian\cite{dicke54} is:
\begin{eqnarray}
  \label{eq:1}
    H &=&
    \sum_{j=1}^{j=nA} 2 \epsilon_j S_j^z
    +
    \sum_{k=2\pi l/\sqrt{A}} \hbar \omega_k \psi^{\dagger}_k \psi^{}_k
    \nonumber\\
    &+&
    \frac{g}{\sqrt{A}} \sum_{j,k} \left(
      e^{2\pi i \vec k \cdot \vec r_j} \psi_k S_j^+ +
      e^{-2\pi i \vec k \cdot \vec r_j} \psi_k^{\dagger} S_j^-
    \right).
\end{eqnarray}
Here $A\rightarrow\infty$ is the quantisation area and $n$ the areal
density of two-level systems, {\it i.e.}\ sites where an exciton may exist.
Without inhomogeneous broadening, the energy of a bound exciton is
$2\epsilon=\hbar\omega_0 - \Delta$, defining the detuning $\Delta$ between
the exciton and the photon.  
When later an inhomogeneously broadened band of exciton energies is
introduced, $\Delta$ will represent the detuning between photon and
centre of the exciton band.
\label{sec:model-parameters}
The photon dispersion, for photons in an ideal 2D cavity of width $w$,
and relative permittivity $\varepsilon_r$ is
\begin{equation}
  \label{eq:2}
  \hbar \omega_k 
  = 
  \hbar \frac{c}{\sqrt{\varepsilon_r}}
  \sqrt{k^2 + \left( \frac{2\pi}{w} \right)^2 }
  \approx
  \hbar \omega_0 + \frac{\hbar^2 k^2}{2m},
\end{equation}
so the photon mass is $m=(\hbar \sqrt{\varepsilon_r}/c) (2\pi/w)$.

The coupling constant, $g$, written in the dipole gauge is,
\begin{equation}
  \label{eq:3}
  g = d_{ab} \sqrt{%
    \frac{e^2}{2 \varepsilon_0 \varepsilon_r} 
    \frac{\hbar \omega_k}{w}
  },
\end{equation}
where $d_{ab}$ is the dipole matrix element.
For small photon wavevectors (w.r.t.\ $1/w$), it is justified to
neglect the $k$ dependence of $g$, {\it i.e.}\ to take
$\omega_k=\omega_0$.
The factor of $1/\sqrt{w}$ is due to the quantisation volume for
the electric field.

The grand canonical ensemble, $\tilde{H}=H-\mu N$, allows the
calculation of equilibrium for a fixed total number of excitations, N,
given by;
\begin{equation}
  \label{eq:4}
      N=  \sum_{j=1}^{nA} \left(
        S_j^z+ \frac{1}{2}
    \right)
    + \sum_{k=2\pi l/\sqrt{A}} \psi^{\dagger}_k \psi^{}_k.
\end{equation}
We therefore define $\hbar\omtil_k=\hbar\omega_k-\mu$ and
$\epstil=\epsilon-\mu/2$.  
Expressing all energies in terms of the scale of the Rabi splitting,
$g\sqrt{n}$, and all lengths via the two-level system density $n$,
there remain only two dimensionless parameters that control the
system, the detuning $\Delta^{\ast}=\Delta/g\sqrt{n}$ and the
photon mass $m^{\ast}=m g / \hbar^2 \sqrt{n}$.
Typical values for current
experiments\cite{deng03:_polar,richard04:_angle_cdte_ii_vi} are
$g\sqrt{n} \approx 10 \mathrm{meV}$, $m \approx 10^{-5}
m_{\mathrm{electron}}$, and taking $n\approx 1/a_{\mathrm{Bohr}}^2
\approx 10^{12} \mathrm{cm}^{-2}$ leads to an estimate of
$m^{\ast}\approx 10^{-3}$.

This model is similar to that studied by Hepp and
Lieb~\cite{hepp73:_equil}.
They considered the case $\mu=0$, {\it i.e.}\ without a bath to fix
the total number of excitations.
The transition studied by Hepp and Lieb was later shown by
Rz\c{a}\.{z}ewski {\it et al.}~\cite{rzazewski75:_prl} to be an
artefact of neglecting $A^2$ terms in the coupling to matter.
The transition in our model, when $\mu\ne0$ does not suffer the same
fate\cite{eastham01:_bose}:
The sum rule of Rz\c{a}\.{z}ewski {\it et al.}\ which prevents condensation no
longer holds when $\mu\ne0$.

In order to integrate over the two-level systems, it is convenient to
represent them as fermions, $S^z=\frac{1}{2}(b^{\dagger}b-a^{\dagger}a)$ and
$S^+=b^{\dagger}a$.  For each site there then exist four states, the
two singly occupied states, $a^{\dagger} \left| 0 \right>$,
$b^{\dagger} \left| 0 \right>$, and the unphysical states $\left| 0
\right>$ and $a^{\dagger} b^{\dagger} \left| 0 \right>$.

Following Popov and Fedotov\cite{popov88:_semion} the
sum over states may be restricted to the physical states by inserting
a phase factor $e^{i(\pi/2)(b^{\dagger}b+a^{\dagger}a)}$.  Since the
Hamiltonian has identical expectations for the two unphysical states, this
phase factor causes the contribution of zero occupied and doubly
occupied sites to cancel, so the partition sum includes
only physical states.  Such a phase factor may then be incorporated as
a shift of the Matsubara frequencies for the fermion fields,
using instead $\omega_n = (n + 3/4) 2\pi T$.
Thus, from here we shall describe the two-level systems as fermions.

\subsection{High energy properties, ultraviolet divergences}
\label{sec:high-energy-prop}

The Hamiltonian~(\ref{eq:1}) is a low energy effective theory, and
will fail at high energies.  
This theory is not renormalisable; there exist infinitely many
divergent one particle irreducible diagrams, and so would need
infinitely many renormalisation conditions.
These divergent diagrams lead to divergent expressions for the free
energy and density.

To treat this correctly, it would be necessary to restore high
energy degrees of freedom, which lead to a renormalisable
theory.
Integrating over such high energy degrees of freedom will recover,
for low energies,  the theory of eq.~(\ref{eq:1}).  
One may then calculate the free energy for the full theory, with
appropriate counter terms.
The low energy contributions will be the same as before, but 
the high energy parts differ, however such high energy
parts are not relevant at low temperatures.
As we are interested only in the low energy properties of this theory,
we will introduce a cut off $K_{\mathrm{m}}$.  The coupling $g$ is
assumed to be zero between excitons and those photons with
$k>K_{\mathrm{m}}$.  

A number of candidates for this cutoff exist; the reflectivity
bandwidth of the cavity mirrors, the Bohr radius of an exciton (where
the dipole approximation fails), and the momentum at which photon
energy is comparable to higher energy excitonic states (for which the
two-level approximation fails).
Which of these scales becomes relevant first depends on the exact
system, however changes in $K_{\mathrm{m}}$ will lead only to 
logarithmic errors in the density.

\section{Mean field and fluctuation spectrum}
\label{sec:greens-funct-fluct}

\subsection{Summary of Mean Field results}
\label{sec:summary-mean-field}

We first briefly present the results of
refs.~\onlinecite{eastham00:_bose,eastham01:_bose} for the mean-field
analysis.  Integrating over the fermion fields yields an effective
action for photons:
\begin{eqnarray}
  \label{eq:5}
  S[\psi]&=&
  \int_0^{\beta} d\tau
  \sum_k
  \psi^{\dagger}_k \left(\partial_{\tau} + \hbar \omtil_k\right)  \psi^{}_k
  + N
  \mathrm{Tr} \ln \left(\calM\right)
  \\
  \label{eq:6}
  \calM^{-1} &=&  
  \left(
    \begin{array}{cc}
      \partial_{\tau} + \epstil & 
      \frac{g}{\sqrt{A}}\sum_k e^{2\pi i\mathbf{k}\cdot\mathbf{r_j}} 
      \psi^{\vphantom{\dagger}}_k \\
      \frac{g}{\sqrt{A}}\sum_k e^{-2\pi i\mathbf{k}\cdot\mathbf{r_j}} 
      \psi^{\dagger}_{k}&
      \partial_{\tau} - \epstil 
    \end{array}
  \right)
  \nonumber.
\end{eqnarray}
We then proceed by minimising $S[\psi]$ for a static uniform $\psi$
and expanding around this minimum.
The minimum, $\psi_0$, satisfies the equation:
\begin{equation}
  \label{eq:7}
  \hbar \omtil_0 \psi_0
  = g^2 n
  \frac{\tanh ( \beta E )}{2 E} \psi_0,
  \quad
  E=\sqrt{\epstil^2 + g^2  \frac{|\psi_0|^2}{A}}.
\end{equation}
which describe a mean-field condensate of coupled coherent photons
and exciton polarisation.
The mean-field expectation of the density is given by
\begin{equation}
  \label{eq:95}
  \rho_{\mathrm{M.F.}} 
  = 
  \frac{\left|\psi_0\right|^2}{A} +
  \frac{n}{2} \left[
    1 -  \frac{\epstil}{E} \tanh(\beta E)
  \right].
\end{equation}
Note that the photon field acquires an extensive occupation, we may
define the intensive quantity $\rho_0=|\psi_0|^2/A$: the density of
photons in the condensate.
This corresponds to an electric field strength of $\sqrt{\hbar\omega_0
  \rho_0/ 2\varepsilon_0}$.

For an inhomogeneously broadened band of exciton energies, eq.~(\ref{eq:7})
should be averaged over exciton energies.
In this case condensation will introduce a gap in the excitation
spectrum of single fermions\cite{eastham01:_bose}:
Whereas when uncondensed there may be single particle excitations of
energy arbitrarily close to the chemical potential, now the smallest
excitation energy is $2g|\psi|/\sqrt{A}$.
The existence of this gap is reflected by features of the collective
mode spectrum discussed below.

\subsection{Connection to Dicke Superradiance}
\label{sec:conn-dicke-superr}

It is interesting to compare this mean-field condensed state,
described by eq.~(\ref{eq:7}), to the superradiant state originally
considered by Dicke~\cite{dicke54}.
Dicke considered a Hamiltonian similar to eq.~(\ref{eq:1}), but with a
single photon mode.
Constructing eigenstates $|L,m\rangle$ of the modulus and z component
of the total spin, $\vec{S} = \sum_j \vec{S_j}$, Dicke showed that for
N particles, the state $|N/2, 0\rangle$ has the highest radiation
rate.

Taking the state described by the mean-field condensate,
eq.~(\ref{eq:7}), and considering the limit as
$\psi_0\rightarrow\infty$ and $T\rightarrow 0$, the equilibrium state
may be written as
\begin{eqnarray}
  \label{eq:8}
  \left| \psi_{\mathrm{cond.}} \right\rangle 
  &=&
  \frac{1}{2^{N/2}}
  \prod_j \left(|\uparrow\rangle_j + |\downarrow\rangle_j \right)
  \\
  \label{eq:9}
  &=&
  \frac{1}{2^{N/2}}
  \sum_{p=0}^N  \sqrt{ N \choose p}
  \left| \frac{N}{2}, \frac{N}{2} - p \right>,
\end{eqnarray}
{\it i.e.}\ a binomial distribution of angular momentum states.

As N tends to infinity, this becomes a Gaussian centred on the Dicke
super-radiant state, with a width that scales like $\sqrt{N}$.
However, it should be noted that the above state represents only the
exciton part; this should be multiplied by a photon coherent state.
In the self consistent state, there exists a sum of terms with
different divisions of excitation number between the photons and
excitons.
These different divisions have a fixed phase relationship.
Such a statement would remain true even if projected onto a state of
fixed total excitation number.
This is different from the Dicke superradiant state, which has no
photon part.
In the Dicke state, the only important coherence is that between the
different ways of distributing excitations between the two-level
systems.

\subsection{Comparison of Mean-Field results with other boson-fermion systems}
\label{sec:comp-mean-field}

The mean-field equations~(\ref{eq:7}),~(\ref{eq:95}) have a form that is
common to a wide range of fermion systems.
However, the form of the density of states, and the nature of fermion
interactions can significantly alter the form of the mean-field phase
boundary.
It is of interest to compare our system to other systems in which
BCS-BEC crossover has been considered, and to note that even at
the mean-field level, important differences emerge.
To this end, we compare the mean-field equations for our modified
Dicke model to the Holland-Timmermans Hamiltonian
\cite{holland01:_reson_super_quant_degen_fermi_gas,timmermans99:_feshb_bose_einst,ohashi03:_super_fermi_feshb},
and to BCS superconductivity\cite{schrieffer83:_theor_super}.

For a gas of fermionic atoms, or BCS, the density of states is
non-zero for all energies greater than zero, whereas for our localised
excitons, without inhomogeneous broadening, it is a $\delta$ function.
One immediate consequence is a difference of interpretation of the
mean-field equations; the number equation and the self consistent
condition for the anomalous Green's function.
For a weakly interacting fermion system the number equation alone
fixes the chemical
potential\cite{randeria:_cross}, and the
self-consistent condition can then be solved to find the critical
temperature.
For our localised fermions, the chemical potential lies below the band
of fermions, and so the density is controlled by the tail of the Fermi
distribution, so temperature and chemical potential cannot be so
neatly separated.
This can remain true even in the presence of inhomogeneous broadening;
the majority of density may come from regions of large density of
states in the tail of the Fermi distribution.

Such differences are also reflected in the density dependence of the
mean-field transition.
The absence of a direct four-fermion interaction means that the
effective interaction strength is entirely due to photon mediated
interactions.
Since our model has the photons at chemical equilibrium with the
excitons,  this effective interaction strength depends strongly
upon the chemical potential.

For BCS superconductors, and the BCS limit for weakly interacting
fermionic atoms, the dependence of critical temperature upon density
is due to the changing density of states, which appears in the self
consistent equation as a pre-factor of the logarithm in the BCS
equation
\begin{math}
  1/g = \rho_{s} \ln(\omega_{D}/T_c)
\end{math}.
%
In contrast, the density dependence of $T_c$ for the Dicke model is
due to the changing coupling strength and occupation of two-level
systems with changing chemical potential.
Such a change of coupling strength with chemical potential also occurs
in the Holland-Timmermans model, near Feshbach resonance, where the
boson mediated interaction is not dominated by the direct four-fermion 
term.
Even with inhomogeneous broadening of energies, unless the chemical
potential remains fixed at low densities, there will be a strong
density dependence of $T_c$ as the effective interaction strength
changes.
The resultant phase boundary with inhomogeneous broadening, at low
densities, is given in section~\ref{sec:effects-inhom-broad}.


\subsection{Effective action for fluctuations}
\label{sec:effect-acti-fluct}

Including fluctuations about the saddle point, $\psi=\psi_0 +
\delta\psi$, one may  write the two-level system inverse
Green's function as $\calM^{-1}=\calM_0^{-1} + \delta\calM^{-1}$,
where
\begin{equation}
  \label{eq:87}
  \delta\calM^{-1} = \frac{g}{\sqrt{A}}
    \sum_k
    \left(
      \begin{array}{cc}
        0 & e^{2\pi i \vec{k}\cdot\vec{r}_j} \delta \psi_k \\
        e^{-2\pi i \vec{k}\cdot\vec{r}_j} \delta \bar{\psi}_k & 0
      \end{array}
    \right),
\end{equation}
thus eq.~(\ref{eq:5}) can be expanded to quadratic order as

\begin{widetext}
  \begin{eqnarray}
    \label{eq:10}
    S
    &=& S[\psi_0] +
    \beta \sum_{i\omega}\sum_k
    (i\omega + \hbar\omtil_k) |\delta\psi_{\omega,k}|^2 
    - \frac{N}{2} \mathrm{Tr} \left(
      \calM_0 \delta\calM^{-1}\calM_0 \delta\calM^{-1}
    \right)
    \\
    \label{eq:11}
    &=&
    \label{eq:12}
    S[\psi_0] +
    \frac{\beta}{2} \sum_{i\omega}\sum_k
    \left( \begin{array}{ll}
        \delta\bar{\psi}_{\omega,k}, &
        \delta\psi_{-\omega,-k} 
      \end{array}
    \right)
    \left( \begin{array}{cc}
        i\omega + \hbar \omtil_k + K_1(\omega) &
        K_2(\omega)\\
        K_2^{\ast}(\omega)&
        -i\omega + \hbar \omtil_k +K_1^{\ast}(\omega)
      \end{array}
    \right)
    \left( \begin{array}{l}
        \delta\psi_{\omega,k} \\
        \delta\bar{\psi}_{-\omega,-k} 
      \end{array}
    \right).
  \end{eqnarray}
\end{widetext}
The matrix between the photon fields can be identified as the inverse
of the fluctuation photon Green's function, $\mathcal{G}^{-1}$, where
the exciton contribution to the quadratic term
(with $\nu=(n + 3/4) 2\pi T$) is
\begin{eqnarray}
  \label{eq:13}
  K_1(\omega)&=&
  \frac{g^2}{A}
  \sum_j
  \sum_{\nu}
  \frac{%
    \left( i\nu + \epstil \right)
    \left( i\nu + i\omega - \epstil \right)
  }{%
    \left( \nu^2 + E^2 \right)
    \left( (\nu+\omega)^2 + E^2 \right)    
  }
  \nonumber\\
  &=&
  g^2n \frac{\tanh(\beta E)}{E}
  \left(
    \frac{
      i \tilde{\epsilon} \omega - 
      E^2 - \tilde{\epsilon}^2 
    }{
      \omega^2  + 4 E^2
    }
  \right)
  - |\alpha| \delta_{\omega},
  \\
  \label{eq:36}
  K_2(\omega)&=&
  \frac{g^2}{A}
  \sum_j
  \sum_{\nu}
  \frac{g^2 \psi_0^2/A}{%
    \left( \nu^2 + E^2 \right)
    \left( (\nu+\omega)^2 + E^2 \right)    
  }
  \nonumber\\
  &=&
  g^2n \frac{\tanh(\beta E)}{E}
  \left(
    \frac{
      g^2 \psi_0^2 /A
    }{
      \omega^2  + 4 E^2
    }
  \right)
  -
  \alpha \delta_{\omega},
  \\
  \label{eq:93}
  \alpha&=&
  g^2 n \beta  \frac{\sech^2(\beta E)}{4 E^2} 
  g^2 \frac{\psi_0^2}{A},
\end{eqnarray}
where the sum over sites assumed no inhomogeneous broadening, and
$\omega$ is a bosonic Matsubara frequency, $2\pi n T$.
The mean-field action, $S[\psi_0]$, is given by:
\begin{equation}
  \label{eq:14}
  S[\psi_0] 
  =
  \hbar \omtil_k |\psi_0|^2
  - 
  \frac{\mu N}{2} 
  - 
  \frac{N}{\beta} \ln\left[\cosh \left(\beta E \right) \right].
\end{equation}

The terms $\alpha\delta_\omega$ occur when the sum over fermionic
frequencies in eqs.~(\ref{eq:13})(\ref{eq:36}) have second order
poles.  These terms must be included in the thermodynamic Green's
function at $\omega=0$.
However, they do not survive analytic continuation, and so do not
appear in the retarded Green's function or in the excitation
spectrum.
This is discussed in Appendix~\ref{sec:lehm-repr-delt}.

Even considering inhomogeneous broadening of exciton energies, such
terms remain as $\delta_{\omega}$, rather than some broadened peak.
This can be understood by considering which transitions contribute to
the excitons' response to a photon, {\it i.e.}\  between which exciton states
there is a matrix element due to the photon.
If uncondensed, the photon couples to transitions between the exciton's
two energy states, $\pm \epsilon$.
In the presence of a coherent field, these energy states mix.
The photon then couples both to transitions $E \rightarrow - E$ and
also to the degenerate transition $E \rightarrow E$.
Since this transition is between the two levels on a single site,
inhomogeneous broadening does not soften the $\delta_{\omega}$ term.

This conclusion differs for models with transitions between
two bands of fermion states.
If transitions are allowed between any pair of lower and upper
band states,  the degenerate transition above is replaced by
a continuum of intra-band transitions.
In our model all intra-band excitations are of zero energy.
If there exists a range of low energy intra-band transitions, these
allow the Goldstone mode to decay, giving rise to Landau
damping~\cite{ohashi97,ohashi03:_super_fermi_feshb}.
For our model, as there is no continuum of modes, no such damping
occurs.

In order to consider fluctuations for an inhomogeneously broadened
system of excitons, a correct treatment requires calculating for a
given realisation of disorder, and then averaging the final results.
Because the position of an exciton matters in its coupling to light,
it would be necessary to include averaging over disorder in both
energy and position of excitons.
However, for low energy modes, we may make an approximation, and
average the expressions for $K_1,K_2$ over exciton energies.

This approximation is equivalent to the assumption that the energies
and positions of excitons sampled by photons of different momenta are
independent and uncorrelated.
Such an approximation evidently cannot hold for high momenta, as
otherwise the number of random variables would become greater than
the number of excitons.
This approximation will also necessarily neglect scattering between
polariton momenta states.
However, such effects involve high energy states (since they require
momenta on the order of the inverse exciton spacing), and can be
neglected in discussing the low energy behaviour.

\subsection{Fluctuation Spectrum}
\label{sec:poles-greens-funct}

Inverting the matrix in the effective action for fluctuations,
eq.~(\ref{eq:12}), one finds the fluctuation Green's function.
The location of the poles of this Green's function give the spectrum
of those excitations which can be created by injecting photons,
measured relative to the chemical potential.

These poles come from the denominator:
\begin{eqnarray}
  \label{eq:16}
  \det\left(\mathcal{G}^{-1}\right)
  &=&
  \left|i\omega + \hbar \omtil_k + K_1(\omega)\right|^2 - 
  \left|K_2(\omega)\right|^2
  \nonumber\\
  &=&
  \frac{
    \left( \omega^2 + \xi_1^2 \right)
    \left( \omega^2 + \xi_2^2 \right)
  }{
    \left( \omega^2 + 4 E^2 \right)
  },
\end{eqnarray}
where, as discussed above, we have ignored the $\delta_{\omega}$
terms, and assumed no inhomogeneous broadening of excitons

In the condensed state the poles are:
\begin{equation}
  \label{eq:17}
  \xi_{1,2}^2 = \frac{1}{2} \left\{
    A(k) \pm 
    \sqrt{A(k)^2 - B(k)
    }
  \right\},
\end{equation}
where
\begin{eqnarray}
  \label{eq:18}
  A(k)&=&4E^2 + (\hbar \omtil_k)^2 + 4\epstil \hbar \omtil_0,
  \\
  \label{eq:19}
  B(k)&=&16 \frac{\hbar^2 k^2}{2m} \left(
    E^2 \hbar \omtil_k - \epstil^2 \hbar \omtil_0
  \right).
\end{eqnarray}

\begin{figure}[htbp]
  \centering
  \includegraphics[width=3.4in]{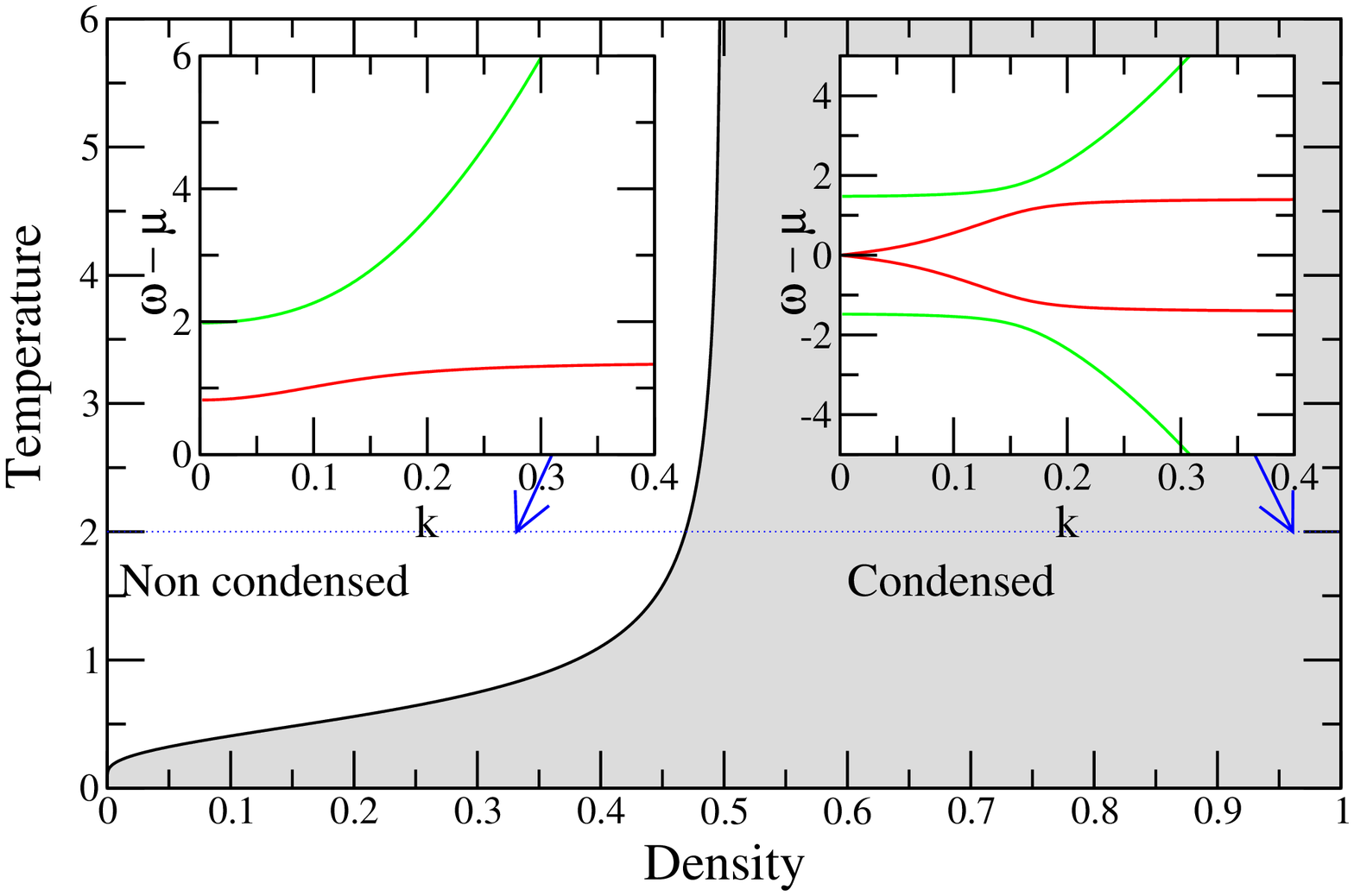}
  \includegraphics[width=3.4in]{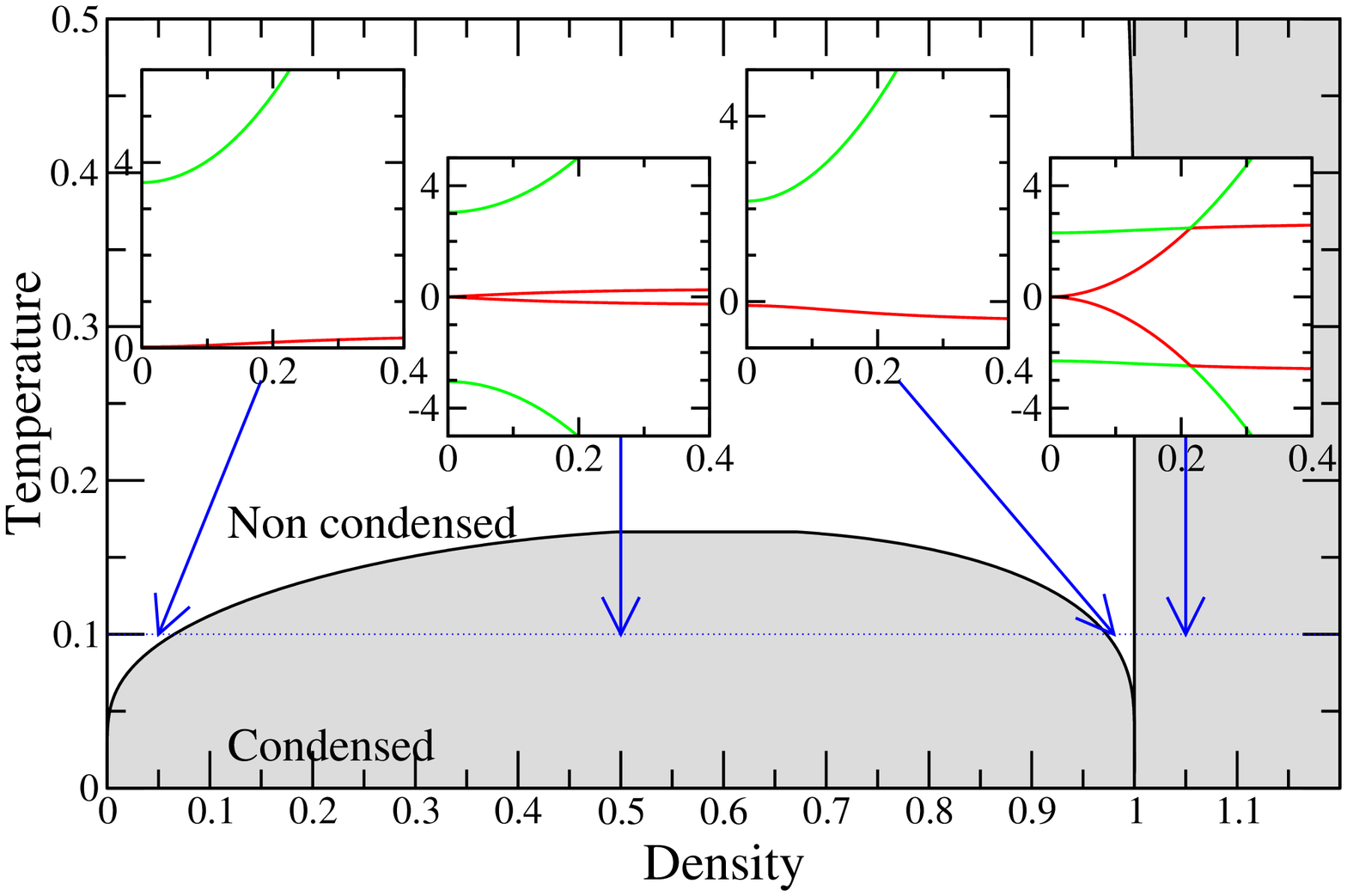}
  \caption{%
    Excitation spectra in the condensed (grey) and uncondensed
    states, superimposed on the mean-field phase diagram, to show
    choice of density and temperature.
    The upper figure is for the photon and exciton resonant, and the
    two spectra are for $T=2 g\sqrt{n}$ and $\mu/g\sqrt{n}=-1.4$ and
    $-0.24$.
    The lower has the exciton detuned by $3g\sqrt{n}$ below the
    photon, and the spectra are for $T=0.1 g\sqrt{n}$ and
    $\mu/g\sqrt{n}=-3.29$, $-3.01$, $-2.54$ and $-0.37$.
    The photon mass is $m^{\ast}=0.01$.
    Temperatures and energies plotted in units of $g\sqrt{n}$,
    densities in units of $n$, and wavevectors in units of $\sqrt{n}$.
  }
  \label{fig:simple_spectrum}
\end{figure}

In the normal state this simplifies further, $K_2$ is zero,
and eq.~(\ref{eq:16}) is replaced by
\begin{equation}
  \label{eq:20}
  \left| i\omega + \hbar \omtil_k + K_1(\omega) \right|^2
  = 
  \left|
    \frac{(i\omega + E_{+})(i\omega + E_{-}) }{(i\omega+2\epstil) }
  \right|^2.
\end{equation}
There are two poles, the upper and lower polariton;
\begin{equation}
  \label{eq:21}
    E_{\pm}=
  \frac{1}{2}
  \left(
    (\hbar \omtil_k + 2 \epstil)
    \pm
    \sqrt{%
      (\hbar \omtil_k - 2 \epstil)^2
      + 4 g^2 n \tanh(\beta\epstil)
    }
  \right).
\end{equation}
The polariton dispersion~(\ref{eq:21}) from localised excitons has the
same structure as from propagating excitons, since the photon
dispersion dominates.
The spectra, both condensed and uncondensed, are shown in
figure~\ref{fig:simple_spectrum}.

The difference between condensed and uncondensed spectra is dramatic:
two new poles appear. 
These arise because the off diagonal terms in eq.~(\ref{eq:12}) mix
photon creation and annihilation operators.
Such a spectrum may be observed in polariton condensation 
experimentally, as one may probe the response to inserting a real photon,
and observing its emission at a later time.  
In the presence of a condensate, the polariton is not a quasi-particle.
Creating a photon corresponds to a superposition of creating and
destroying quasi-particles.
At non-zero temperature, a population of quasi-particles exists, so
processes where a quasi-particle is destroyed is possible.
This means that a process in which a photon, with energy less than the
chemical potential, is added to the system is possible ({\it i.e.}\
$P_{\mathrm{absorb}}$ as defined in eq.~(\ref{eq:86})).
However, if the system is experimentally probed with photons of energy
less than the chemical potential, what will be observed is gain, since
the absorption of photons is more than cancelled by spontaneous and
stimulated emission.

At small momentum, $\xi_1$ corresponds to phase fluctuations
of the condensate, {\it i.e.}\ it is the Goldstone mode, and
has the form $\xi_1=\pm \hbar c k$, with,
\begin{equation}
  \label{eq:22}
    c= \sqrt{%
    \frac{1}{2m}
    \left(\frac{4\hbar \omtil_0 g^2 n}{\xi_2(0)^2}\right)
    \left(\frac{|\psi_0|^2}{N}\right)}
  \approx \sqrt{\frac{\lambda}{2m}\frac{\rho_0}{n}}.
\end{equation}
The second expression is, for comparison, the form of the Bogoliubov
mode in a dilute Bose gas, of interaction strength $\lambda$.
As $\psi_0$ increases, the phase velocity first increases, then decreases.
The decrease is due to the saturation of the effective exciton-photon
interaction.

The leading order corrections $\xi_1=\pm c k + \alpha k^2$ are of interest
for considering Beliaev decay of phonons\cite{beliaev58}.
If $\alpha<0$, kinematic constraints prevent the decay of phonons.
For the Bogoliubov spectrum in a dilute Bose gas, $\alpha=0$, and the
cubic term becomes relevant.  Here, both signs are possible; according
to whether the spectrum crosses over to the quadratic lower polariton
dispersion before this crosses over to a flat exciton dispersion.
In most cases, $\alpha>0$ and $\xi_1$ will have a point of inflection,
but if the exciton is detuned below the photon, then for certain
densities, $\alpha<0$, and the curvature is always negative.

The modes may also be compared to the Cooperon modes in BCS theory.
At small momenta, although the mode $\xi_1$ becomes a pure
phase fluctuation, the other mode $\xi_2$ is not an amplitude
fluctuation.
To see why this is so, it is helpful to
rewrite the action in equation~(\ref{eq:12}) in terms of the Fourier
components of the transverse and longitudinal fluctuations of the
photon field.
These components, at quadratic order, are equivalent to the phase and
amplitude components, and are given by:
\begin{eqnarray}
  \label{eq:23}
  \delta{\psi}_{\omega,k} &=& \psi_L(\omega,k) + i\psi_T(\omega, k),\nonumber\\
  \delta\bar{\psi}_{\omega,k} &=& \psi_L(-\omega,-k) - i \psi_T(-\omega, -k).
\end{eqnarray}
\begin{widetext}
In terms of these new variables, the action may be written:
\begin{equation}
  \label{eq:26}
  S=S[\psi_0] +
  \frac{\beta}{2} \sum_{i\omega,k}
  \left( \begin{array}{ll}
      \psi_L(-\omega,-k) & 
      \psi_T(-\omega, -k)
    \end{array}
  \right)
  \left( \begin{array}{cc}
      \hbar \omtil_k + \Re(K_1) + K_2 & 
      - \omega - \Im(K_1)  \\
       \omega + \Im(K_1) &
      \hbar \omtil_k + \Re(K_1) - K_2 
    \end{array}
  \right)
  \left( \begin{array}{l}
      \psi_L(\omega,k) \\
      \psi_T(\omega,k)
    \end{array}
  \right).
\end{equation}
\end{widetext}

Due to the off-diagonal components, the eigenstates are mixed amplitude
and phase modes.
This mixing vanishes only where $\omega$ is small, which means that
the lowest energy parts of the Goldstone mode are purely phase
fluctuations.
Since the amplitude mode at $k=0$ has a non-zero frequency, it will
mix with the phase mode.

The off diagonal terms come from two sources.
The dynamic photon field leads to the term $\omega$.
The fermion mediated term $\Im(K_1)$ will be non zero if the density
of states is asymmetric about the chemical potential.
For the Dicke model, both terms contribute to mixing since:
\begin{equation}
  \label{eq:27}
  \Im(K_1) =   g^2n \frac{\tanh(\beta E)}{E}
  \left(
    \frac{%
      \tilde{\epsilon} \omega
    }{%
      \omega^2  + 4 E^2
    }
  \right).
\end{equation}
Unless $\epstil=0$, the density of states is not symmetric about
the chemical potential, and so this term is non-zero.

For BCS superconductivity the pairing field is not dynamic, so there
is no off diagonal $\omega$ term.
However there can still be mixing due to $\Im(K_1)$, which is given
by:
\begin{equation}
  \label{eq:28}
  \Im(K_1) = \sum_{i\nu, q} 
  \frac{%
    \nu (\epsilon_{q} - \epsilon_{q+Q}) + \omega \epsilon_q
  }{%
    (\nu^2 + \epsilon_q^2 + \Delta^2)
    ((\nu+\omega)^2 + \epsilon_{q+Q}^2 + \Delta^2)
  },
\end{equation}
in which $\nu$ is a fermionic Matsubara frequency, $(2n+1)\pi T$, and
$\Delta$ the superconducting gap.
Note that this coupling now depends on the momentum transfer $Q$ as well
as energy $\omega$.
If the density of states is symmetric, {\it e.g.}\ $\epsilon_q=v_F q$,
then at $Q=0$, $\Im(K_1)$ will be zero, and as the boson field has no
dynamics, the modes will then entirely decouple.
In real superconductors, symmetry of the density of states about the
chemical potential is only approximate, and so some degree of mixing
will occur.

\subsection{Luminescence spectrum}
\label{sec:luminesence-spectrum}

Although four poles exist, they may have very different
spectral weights. 
At high momentum, the weights of all poles except the highest vanish,
and the remaining pole follows the bare photon dispersion.

\begin{figure}[htbp]
  \centering
  \includegraphics[width=3.3in]{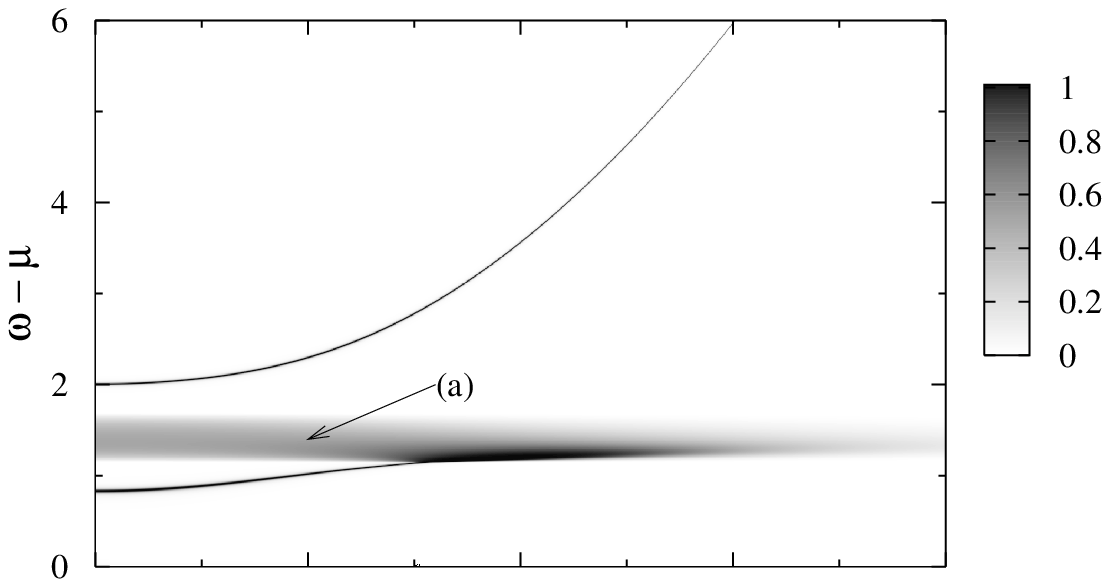}
  \includegraphics[width=3.3in]{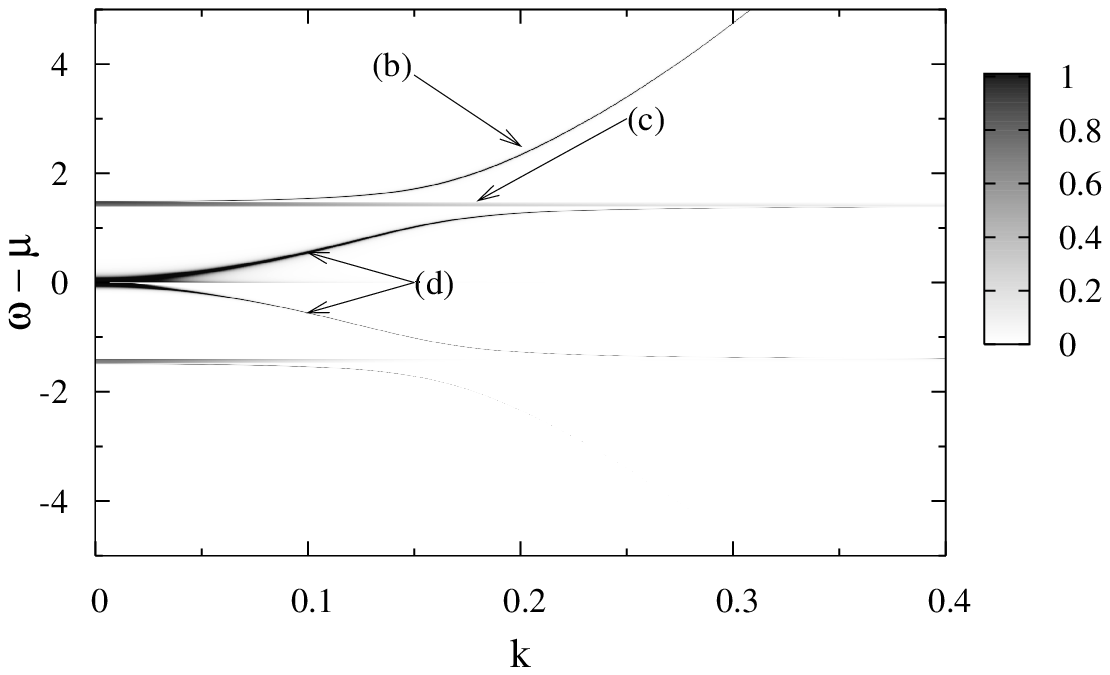}
  \vspace{2ex}
  \includegraphics[width=3.3in]{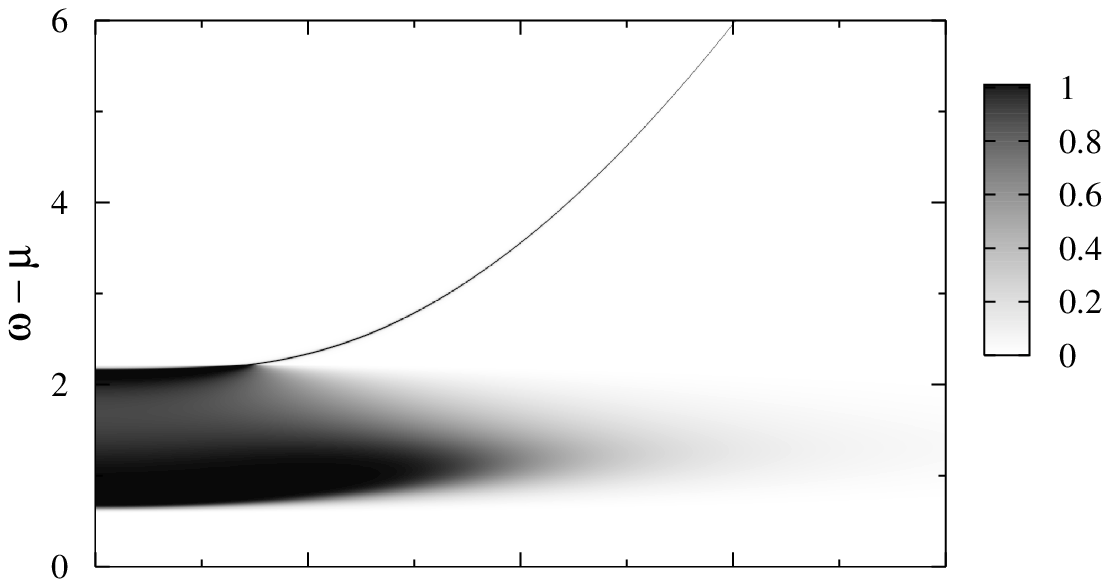}
  \includegraphics[width=3.3in]{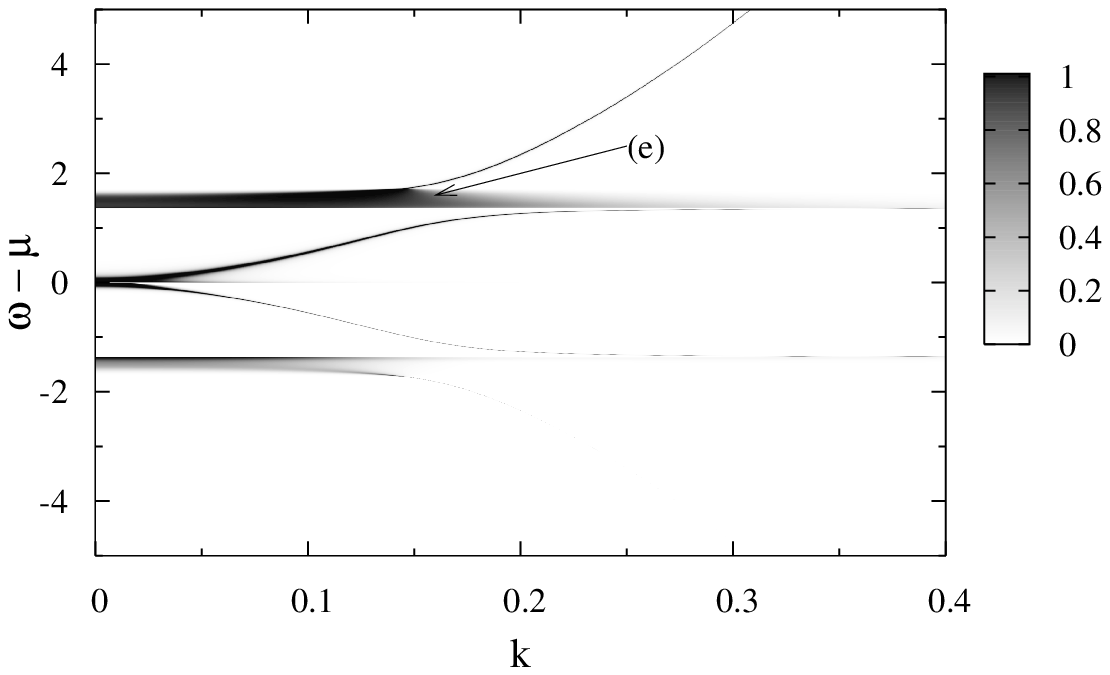}
  \caption{%
    Incoherent luminescence vs\ momentum (x axis) and energy (y
    axis) calculated above (panels 1 and 3) and below (panels 2 and 4)
    phase transition.
    Panels 1 and 2 are for inhomogeneous broadening of $0.1g\sqrt{n}$,
    and 3 and 4 are for $0.3g\sqrt{n}$.
    All other parameters are the same as in the unbroadened case
    shown in panel 1 of figure~\ref{fig:simple_spectrum}.
    In order to map the infinite range of intensities to a finite
    scale, the colour is allocated as the hyperbolic tangent of
    intensity.
    Energies are in units of $g\sqrt{n}$, and wavevectors in units
    of $\sqrt{n}$.
  }
  \label{fig:dos}
\end{figure}

Such effects can be seen more clearly by plotting the incoherent
luminescence spectrum.
As discussed in appendix~\ref{sec:lehm-repr-delt}, this can be found
from the Green's function as
\begin{eqnarray}
  \label{eq:90}
  P_{\mathrm{emit}}(x)
  &=&
  2 n_{\mathrm{B}}(x)
  \Im\left[\mathcal{G}_{00}(i\omega=x+i\eta)\right].  
\end{eqnarray}
It is easier to observe how the spectral weight associated with poles
changes after adding inhomogeneous broadening. 
Figure~\ref{fig:dos} plots the luminescence spectrum for the same
parameters as in figure~\ref{fig:simple_spectrum}, but with
inhomogeneous broadening.
This will broaden the poles, except for the Goldstone mode (labelled
(d) in figure~\ref{fig:dos}), as discussed above.
The distribution of energies used is, for numerical efficiency, a
cubic approximation to a Gaussian, with standard deviations
$0.1g\sqrt{n}$ and $0.3g\sqrt{n}$.
The results with a Gaussian density of states have been compared to
this cubic approximation, and no significant differences occur.

In the condensed case, a third pair of lines are visible, (labelled
(c) in figure~\ref{fig:dos}).
These correspond to the minimum energy for a neutral excitation,
$2g\sqrt{n} \psi_0$ , {\it i.e.}\  flipping the spin on a single site.
They are analogous to a particle-hole excitation in BCS, which have
twice the energy of the gap for adding a single particle.
This energy is the smallest ``inter-band excitation''.
If the inhomogeneous broadening is small compared to the gap then
these extra lines will be less visible, as seen in panel 2.
For large inhomogeneous broadening , the peak at the edge of the gap
(c) and the upper polariton (b) will merge, as shown by (e) in panel
4.
This will result in a band of incoherent luminescence separated from
the coherent emission at the chemical potential by the gap.
Such structure, although associated with the internal structure of a
polariton, can be seen even at densities where the transition
temperature is adequately described by a model of structureless
polaritons.

It is also interesting to consider the uncondensed cases.
In panel 1, with the smaller broadening, three lines are visible; the
upper and lower polariton, and between them luminescence from excitons
weakly coupled to light (labelled (a)).
These weakly coupled states arise from excitons further away from
resonance with the photon band.
Although there may be a large density of such states, they make a much
smaller contribution to luminescence than the polariton states, because
of their small photon component.
With larger broadening, these weakly coupled excitons form a continuum
stretching between the two modes, as shown in panel 3.

\subsection{Momentum distribution of photons}
\label{sec:moment-distr-phot}

From the Green's function for photon fluctuations, one can calculate
the momentum distribution of photons in the cavity. 
For a two-dimensional cavity coupled via the mirrors to
three-dimensional photons outside, the momentum distribution may be
observed experimentally from the angular
distribution\cite{deng03:_polar}.

When uncondensed, $N(p)$ is given by
\begin{eqnarray}
  \label{eq:29}
  N(p)
  &=&
  \lim_{\delta\rightarrow0^+}
  \left< \psi^{\dagger}_p(\tau + \delta) \psi^{}_p(\tau) \right> 
  \nonumber\\
  &=&
  \lim_{\delta\rightarrow 0^+}
  \beta \oint \frac{dx}{2\pi i} n_B(z) e^{\delta z} 
  {\mathcal G}_{11}(iz,p).
\end{eqnarray}

However when condensed, since the system is two-dimensional, it
is necessary to treat fluctuations more carefully.
Writing $\psi(r)=\sqrt{\rho_0 + \pi(r)} e^{i\phi(r)}$, the action
involves only derivatives of the phase, showing that large phase
fluctuations are possible.
For quadratic fluctuations, one can use the
matrix in eq.~(\ref{eq:26}) describing transverse and longitudinal
modes, and relate these to the phase and amplitude excitations.

At low enough temperatures, it is possible to calculate $N(p)$ by
considering only the phase mode\cite{keeling04:_angul}.
Thus, neglecting amplitude fluctuations gives;
\begin{eqnarray}
  \label{eq:30}
  N(p)
  &=&
  \frac{1}{A} \int d^2r \int d^2r^{\prime}
  e^{i\vec{k}\cdot\left(\vec{r} - \vec{r}^{\prime}\right) }
  \rho_0
  \left<
    e^{i \left( \phi(\vec{r}) - \phi(\vec{r}^{\prime}) \right) }
  \right>
  \nonumber\\
  &=&
  \rho_0 \frac{1}{A} \int d^2 R \int d^2 t
  e^{i\vec{k}\cdot \vec{t} }
  e^{- D(\vec{R}+\vec{t}/2, \vec{R}-\vec{t}/2) },
\end{eqnarray}
where $\rho_0$ is taken as the mean-field photon density.
The phase correlator, found by inverting the amplitude/phase action, is:
\begin{eqnarray}
  \label{eq:31}
  D(\vec{r}, \vec{r}^{\prime}) 
  &=& 
  \int  \frac{d^2 k}{(2\pi)^2}
  \left[1-\cos\left(\vec{k}.(\vec{r}-\vec{r}^{\prime})\right) \right]
  \frac{m}{\beta \rho_0 \hbar^2 k^2}
  \nonumber\\
  &\approx&
  \frac{m}{2\pi \beta \rho_0 \hbar^2} 
  \ln\left( \frac{ |\vec{r}-\vec{r}^{\prime}|}{\xi_T} \right).
\end{eqnarray}
The thermal length is $\xi_T =\beta c$, where $c$ is the velocity
of the sound mode from eq.~(\ref{eq:22}).  
This comes from the energy scale at which fluctuations become cut by
the thermal distribution.

In this approximation, eq.~(\ref{eq:30}) may be evaluated
exactly\cite{marchetti04:_conden_cavit_polar_disor_envir} giving
\begin{eqnarray}
  \label{eq:32}
  N(p) 
  &=&
  2\pi \rho_0 \frac{(p\xi_T)^\eta}{p^2}
  \int_0^\infty x^{1-\eta} J_0(x) dx
  \nonumber\\
  &=&
  2\pi \rho_0 \frac{\xi_T^\eta}{p^{2-\eta}}
  2^{1-\eta} \frac{\Gamma(1-\eta/2)}{\Gamma(\eta/2)},
\end{eqnarray}
where $\eta=m/2\pi \beta \rho_0 \hbar^2$ controls the power
law decay of correlations.
The second line follows from an identity (see 
ref.~\onlinecite{whittaker27:_moder_analy} exercise XVII.32).
This is valid only for small $\eta$, far from the transition.
The Kosterlitz-Thouless transition\cite{kosterlitz73:_order} occurs
when $\eta$ becomes large, an approximate estimate of the transition
is at $\eta=2$.

\begin{figure}[htbp]
  \centering
  \includegraphics[width=3.4in]{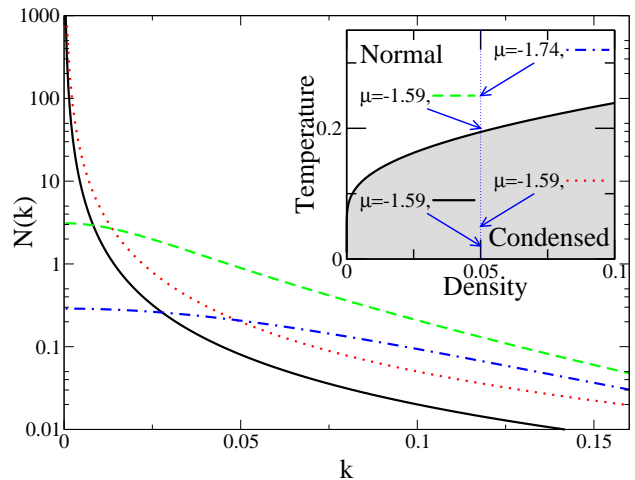}
  \caption{Momentum distribution of photons, from which follows the 
    angular distribution.  These are plotted for low temperature
    condensed systems, where the approximation of including only phase
    fluctuations is valid, and for the simpler, uncondensed case.
    The inset illustrates the choices of density and temperature.
    (Parameters $\Delta^{\ast}=1$, $m^{\ast}=0.01$, wavevector plotted
    in units of $g\sqrt{n}$, temperature in units of $g\sqrt{n}$,
    density in units of $n$, and $N(k)$ in arbitrary units.)
    }
  \label{fig:nofk}
\end{figure}

The momentum distribution can therefore be calculated both at low
temperatures, where such a scheme holds, and at high temperatures,
when uncondensed.
This is shown in figure~\ref{fig:nofk}.
When condensed, the power law divergence leads to a peak normal to the
plane.
This peak reflects coherence between distant parts of the system, so
in a finite system this peak is cut at small momenta\cite{keeling04:_angul}.

\section{Fluctuation correction to the mean-field theory}
\label{sec:fluct-corr-dens}

In this section, we calculate the fluctuation corrections to
the mean-field density, and thus to the mean-field phase boundary.
Our method is similar to that of Nozi\`eres and
Schmitt-Rink\cite{nozieres85:_crossover}, who
studied fluctuation corrections to the BCS mean-field theory for a
model of interacting, propagating fermions.
However, because our model differs from that of Nozi\`eres and
Schmitt-Rink, our approach to including second order fluctuations will
also differ.
We begin by presenting a brief summary of the method used by
Nozi\`eres and Schmitt-Rink, and further developed by
Randeria\cite{randeria:_cross}.
We then discuss how our approach differs from their work.

\subsubsection{Fluctuation corrections in three dimensions}
\label{sec:fluct-corr-three}

To consider fluctuation corrections to a mean field-theory, one first
needs to find the partition function in terms of a coherent state
path integral for a bosonic field,
\begin{displaymath}
  \mathcal{Z}=\int \mathcal{D}\psi \exp(-S[\psi])
\end{displaymath}
For the Dicke model, $S[\psi]$ is given in eq.~(\ref{eq:5}).
In the work of Nozi\'eres and Schmitt-Rink, $S[\psi]$ resulted from
decoupling a four-fermion interaction, and then integrating over the
fermions.
This effective action may be understood as a Ginzburg-Landau theory,
with coefficients that are functions of temperature and chemical
potential as well as the parameters in the Hamiltonian.
To find the mean-field phase boundary, one needs both to find the
values of temperature and chemical potential where the transition
occurs, and to calculate the density evaluated at these parameters.

For the mean-field theory, the action is evaluated for a static
uniform field, $\psi_0$, and minimised w.r.t.\ this field.
At the critical temperature, a second order phase transition occurs,
and the minimum action moves to a non-zero $\psi_0$.
The density is found by differentiating the free energy w.r.t.\ 
chemical potential.
For the mean field theory, the free energy is approximated by
the action evaluated at $\psi_0$.

To go beyond the mean field theory, one can expand the effective
action about the saddle point:
\begin{equation}
  \label{eq:96}
  \mathcal{Z}
  \approx
  e^{-S[\psi_0]}
  \int \mathcal{D} \delta\psi 
  \exp \left(
    - \frac{1}{2} \left.
      \frac{\partial^2 S}{\partial \psi^2}
    \right|_{\psi_0} (\delta\psi)^2
  \right)
\end{equation}
This gives an improved estimate of the free energy, from which follows
an improved estimate of the density, and thus of the phase boundary.
In three dimensions (but not in two, as discussed below), one only
needs an estimate of the density at the mean-field critical
temperature.
It is therefore sufficient to consider an expansion about the normal
state saddle point, $\psi=0$.
Such fluctuations may be understood as the contribution to the density
from non-condensed pairs of particles, whereas the mean-field estimate
of density included only unpaired fermions.
These corrections will increase the density at a fixed critical
temperature, or equivalently decrease the critical temperature for a
given density.

Because of features of our model, our approach differs from
that of Nozi\`eres and Schmitt-Rink; two differences in our model are
of particular importance.
Firstly, our boson field is dynamic, and there exists a chemical
potential for bosons.
In this respect, our model is closer to the boson-fermion models%
\cite{holland01:_reson_super_quant_degen_fermi_gas,%
  timmermans99:_feshb_bose_einst} studied in the context of Feshbach
resonances, for which Ohashi \& Griffin
\cite{ohashi03:_super_fermi_feshb} have studied fluctuation
corrections.
Secondly we consider a two-dimensional system; this requires
calculation of fluctuations in the presence of a condensate, as
discussed in the next section.

\subsubsection{Fluctuations in two dimensions}
\label{sec:outl-dens-calc}

To find the fluctuation-correction to the mean-field phase boundary in
two dimensions, it is necessary to consider fluctuations in the
presence of a condensate.
Considering fluctuations in the normal state would lead to the conclusion
that the normal state can support any density:
As one approaches the mean-field critical temperature, the fluctuation
density will become infra-red divergent, allowing any density.
This correctly indicates that no long-range order exists in two
dimensions; however a Kosterlitz-Thouless
\cite{kosterlitz73:_order,nelson77:_univer_jump_super_densit_two_dimen_super}
transition does occur.

Instead one must start by considering fluctuations in the presence of a
condensate.
This gives a density, defined by the total derivative of free energy
w.r.t.\ chemical potential, of the form:
\begin{equation}
  \label{eq:33}
  \rho=- \frac{d F}{d\mu} = 
  - \frac{\partial F}{\partial \mu} 
  - \frac{\partial F}{\partial \psi_0} \frac{d \psi_0}{d \mu}.
\end{equation}

When considering fluctuation corrections in the presence of a
condensate, in any dimension, one must take care to consider
the depletion of the order parameter due to the interaction between
condensed and uncondensed particles.
This is discussed in detail in section~\ref{sec:weakly-inter-bose}.
Such a depletion means that, for a fixed temperature, the critical
value of the chemical potential changes; at the mean-field critical
chemical potential the formula for total density may become negative.
It is therefore necessary to make a separate estimate of the order
parameter in the presence of fluctuations, and define the phase
boundary where the order parameter goes to zero.

In three dimensions such a calculation can be achieved by identifying
parts of the density as the population of the ground state and of
fluctuations (as discussed below in section~\ref{sec:hugenh-pines-relat}).
In two dimensions, no true condensate exists,  but a quasi-condensate
with a cutoff $k_0$ can be considered.  
As discussed by Popov\cite{popov:_chap6}, the quasi-condensate and
fluctuation densities both contain terms which diverge logarithmically
as $k_0\rightarrow 0$, but these divergences cancel in the total
density.

Instead, in two dimensions, one must consider an alternate definition
for the location of the phase boundary.
Since the transition is a Kosterlitz-Thouless transition, one should
map the problem to the two-dimensional Coulomb
Gas\cite{khawaja02:_low_bose}.
This requires the vortex core energy and strength of vortex-vortex
interactions, which both scale as $\hbar^2 \rho_s / 2 m$, where $\rho_s$
is a superfluid response density.
The phase transition thus occurs when $\rho_s = \# 2 m T/\hbar^2$.
The numerical pre-factor depends on the vortex core structure.
However, approximating the critical condition by $\rho_s=0$ leads to
only a small shift to $T_C$.

\subsection{Total derivatives and negative densities}
\label{sec:weakly-inter-bose}

This section discusses the effects and interpretation of the second
term in eq.~(\ref{eq:33}).
In ref.~\onlinecite{ohashi03:_super_fermi_feshb}, Ohashi \& Griffin
define the density as the partial derivative of free energy w.r.t.\ 
chemical potential,
neglecting the second term in eq.~(\ref{eq:33}), which they describe
as a higher-order correction under the Gaussian fluctuation
approximation.
As shown below in sec.~\ref{sec:gauss-fluc-appr}, the contribution of
the second term in eq.~(\ref{eq:33}) to the density should not
necessarily be neglected in the Gaussian fluctuation approximation.
Section~\ref{sec:total-deriv-negat} shows explicitly that the two
terms in eq.~(\ref{eq:33}) are of the same order.
The existence of the second term is crucial in finding a finite
density in two dimensions, however may be less important in three
dimensions.

\subsubsection{Gaussian fluctuation approximation}
\label{sec:gauss-fluc-appr}

In section~\ref{sec:total-deriv-negat} we will show that the second
term in eq.~(\ref{eq:33}) is of the same order as the first.
Before this, we explain why the second term of eq.~(\ref{eq:33})
should not be automatically neglected.
Even though it is of the form
\begin{math}
  \partial^3 S/\partial \psi_0^3
\end{math}, 
such terms are not necessarily small, and can contribute to the
density at quadratic order.

To see that a Gaussian theory may be correct even if such third order
terms are not small, consider an expansion of the effective action,
\begin{equation}
  \label{eq:34}
  S = S[\psi_0] +
  \frac{d^2S}{d \psi_0^2} \delta\psi^2 + 
  \frac{d^3S}{d \psi_0^3} \delta\psi^3 + \ldots
\end{equation}
A Gaussian approximation is justified if, using this action, the
expectation of the cubic term is less than the quadratic term.
This condition can be written as
\begin{equation}
  \label{eq:35}
  \left( \frac{d^2 S}{d \psi_0^2} \right)^{3/2} 
  \gg 
  \frac{d^3 S}{d \psi_0^3}.
\end{equation}
This need not require that the coefficient of the cubic term is
smaller, {\it i.e.}\ that
\begin{displaymath}
  \frac{d^2 S}{d \psi_0^2}
  \gg 
  \frac{d^3 S}{d \psi_0^3}.
\end{displaymath}
In fact, if both terms are of the same order, but large;
\begin{math}
  d^2 S / d \psi_0^2
  \simeq 
  d^3 S / d \psi_0^3
  \gg 1,
\end{math}
then the condition~(\ref{eq:35}) is fulfilled.

Writing the free energy including fluctuations schematically as
\begin{equation}
  \label{eq:76}
  F = S[\psi_0] + \ln \left[
    \int \mathcal{D} \delta\psi
    \exp \left( - \frac{d^2S}{d \psi_0^2} \delta\psi^2  \right)
  \right],
\end{equation}
the fluctuation contribution to the second term in
eq.~(\ref{eq:33}) will take the form
\begin{equation}
  \label{eq:77}
  \rho = \ldots +
  \left<
    \frac{d^3S}{d \psi_0^3} \delta\psi^2
  \right>
  \frac{d\psi_0}{d\mu},
\end{equation}
where $\langle\ldots\rangle$ signifies averaging over the fluctuation
action.
Thus, in calculating the condensate density, there is a term which
depends on 
\begin{math}
  \partial^3 S/\partial \psi_0^3
\end{math}
but only involves second order expectations of the fields.
Since
\begin{math}
  \partial^3 S/\partial \psi_0^3
\end{math}
is not necessarily small, and it contributes to the density at
quadratic order, there is no a priori argument for neglecting this
term.
In the following we show explicitly that this term should be included.

\subsubsection{Total derivatives for a dilute Bose gas}
\label{sec:total-deriv-negat}

The following discussion will show explicitly that the two terms in
equation~(\ref{eq:33}) are of the same order for a weakly interacting
dilute Bose gas (W.I.D.B.G.),
\begin{equation}
  \label{eq:37}
  H - \mu N =  \sum_k (\epsilon_k - \mu) a^{\dagger}_k a^{}_k +
  \frac{g}{2}\sum_{k,k^{\prime}, q}
  a^{\dagger}_{k+q} a^{\dagger}_{k^{\prime}-q} a^{}_k a^{}_{k^{\prime}}.
\end{equation}
Further, the terms in eq.~(\ref{eq:33}) will be interpreted by
considering the Hugenholtz-Pines relation at one loop order, as
discussed in ref.~\onlinecite{popov:_chap6}.

\paragraph{Saddle point and fluctuations.}
\label{sec:saddle-point-fluct}

To find the free energy, consider the static uniform saddle point,
$\langle a^{\dagger}_0 a^{}_0 \rangle=|A|^2 = \mu/g$, and quadratic
fluctuations, which are governed by the Hamiltonian:
\begin{eqnarray}
  \label{eq:38}
  H_{\mathrm{eff}}
  &=& 
  \sum_k (\epsilon_k - \mu + 2 g A^2) a^{\dagger}_k a^{}_k 
  \nonumber\\ &+&
  \frac{g A^2}{2} \left(
    a^{\dagger}_k  a^{\dagger}_{-k} +   a^{}_k a^{}_{-k}
  \right).
\end{eqnarray}
Thus, by a Bogoliubov transform, the free energy and density become
\begin{eqnarray}
  \label{eq:40}
  F&=& -\mu A^2 + \frac{g}{2}A^4 
  \nonumber\\
  &+& 
  \sum_k \left(
    \frac{1}{\beta} \ln\left( 1- e^{-\beta E_k} \right) +
    \frac{1}{2}(E_k - \epsilon_k - \mu)
  \right),\\
  \label{eq:41}
  \rho &=& 
  A^2 - \sum_k \left(
    n_B(E_k) \frac{\epsilon_k}{E_k}
    + \frac{\epsilon_k - E_k}{2E_k}
  \right),
\end{eqnarray}
where $E_k=\sqrt{\epsilon_k(\epsilon_k+2\mu)}$.

The total density is thus less than the saddle point $A^2$, and could
be negative.
From the form of the fluctuation Hamiltonian, eq.~(\ref{eq:38}), using
$g A^2=\mu$, it can be seen that the fluctuation contribution is:
\begin{equation}
  \label{eq:42}
  \rho_f=
  - \sum_k
  \left\{
    \left< a^{\dagger}_k a^{}_k \right> +
    \frac{1}{2}
    \left(
      \left< a^{\dagger}_k a^{\dagger}_{-k} \right> +
      \left< a^{}_k a^{}_{-k} \right>
    \right)
  \right\}.
\end{equation}
Using the results:
\begin{eqnarray}
  \label{eq:43}
  \left< a^{\dagger}_k a^{}_k \right>
  &=&
  n_B(E_k) \frac{\epsilon_k + \mu}{E_k}
  + \frac{\epsilon_k +\mu - E_k}{2E_k},
  \\
  \label{eq:44}
  \left< a^{\dagger}_k a^{\dagger}_{-k} \right>
  &=&
  \left< a^{\vphantom{\dagger}}_k a^{\vphantom{\dagger}}_{-k} \right>
  =
  -\frac{\mu}{E_k}
  \left(
    n_B(E_k) + \frac{1}{2}
  \right),
\end{eqnarray}
it is clear this matches eq.(\ref{eq:41}).

In contrast, taking partial derivatives, and neglecting the second
term in eq.~(\ref{eq:33}) gives 
\begin{math}
  \rho_f=\sum_k \left< a^{\dagger}_k a^{}_k \right> 
\end{math}.
In two dimensions, for $\mu \ne 0$, this expression will be infra-red
divergent, while eq.~(\ref{eq:41}) is not.

\paragraph{Hugenholtz-Pines relation}
\label{sec:hugenh-pines-relat}

To identify the meaning of the terms in eq.~(\ref{eq:42}), one can
consider the Hugenholtz-Pines relation for the normal and anomalous self
energies $A(\omega,k)$, $B(\omega,k)$ respectively:
\begin{equation}
  \label{eq:45}
  A(0,0)-B(0,0)=\mu,
\end{equation}

The approximations in the previous section are equivalent to
evaluating $A$ and $B$ at one loop order.
As explained by Popov\cite{popov:_chap6}, this becomes
\begin{eqnarray}
  \label{eq:94}
  \mu 
  &=&
  2g(\rho_0+\rho_1) - g(\rho_0 + \tilde{\rho_1}) \nonumber\\
  &-& 
  2 g^2 \rho_0 
  \sum_{k} \bigl(
    \mathcal{G}(k) \mathcal{G}(-k) -   \mathcal{G}_1(k) \mathcal{G}_1(-k)
  \bigr).
\end{eqnarray}
Here $\rho_0$ is the new condensate density, $\rho_1$ the density of
non-condensed particles, and $\tilde{\rho}_1$ is an anomalous particle
density,
\begin{math}
  \tilde{\rho}_1=\sum_k \mathcal{G}_1(k)
\end{math},
with $\mathcal{G}_1$ the anomalous Green's function.
The last term is a second order correction due to the three boson
vertices of the form $g A (a^{\dagger} a^{\dagger} a + a^{\dagger} a
a)$.
This last term in eq.~(\ref{eq:94}) can be evaluated to be
$2g\tilde{\rho_1}$, leading to the result
\begin{equation}
  \label{eq:46}
  \rho_0=\frac{\mu}{g} - (2\rho_1 + \tilde{\rho}_1),
\end{equation}
showing that $\rho_0+\rho_1$ is less than the saddle point density.

Compare this expression for the total density to that from saddle
point and fluctuations, 
\begin{equation}
  \label{eq:78}
  \rho=
  \rho_0+\rho_1 
  = 
  \rho_{\mathrm{s.p.}} 
  - 
  \frac{\partial F_{\mathrm{fluct}}}{\partial \mu}
  -
  \frac{d \psi_0}{d \mu}
  \frac{\partial F_{\mathrm{fluct}}}{\partial \psi_0},
\end{equation}
where $\rho_{\mathrm{s.p.}}=\mu/g$ is the saddle point expectation of
the density, and $F_{\mathrm{fluct}}$ is the free energy from the 
fluctuation contributions.
Since $\rho_1$, the density of non-condensed particles can be identified
as
\begin{equation}
  \label{eq:79}
  \rho_1
  =
  \sum_{k} \left< a^{\dagger}_k a^{\vphantom{\dagger}}_k \right>
  =
  -\frac{\partial F_{\mathrm{fluct}}}{\partial \mu},
\end{equation}
one must identify the depleted condensate density, eq.~(\ref{eq:46}) with
\begin{equation}
  \label{eq:47}
  \rho_0=\frac{\mu}{g} -
  \frac{d \psi_0}{d \mu}
  \frac{\partial F_{\mathrm{fluct}}}{\partial \psi_0}.
\end{equation}

The derivatives w.r.t.\ the order parameter therefore describe a
depletion of the order parameter due to fluctuations.  
Physically, interactions between the condensate and the finite
population of non-condensed particles (at finite temperature) push up
the chemical potential.
With such a theory, there now exists a region of parameter space where
which is not condensed, but $\mu>0$.  
In such a region it is essential to include modifications of the
particle spectrum due to interactions to describe the normal state.

\paragraph{Comparison of methods}
\label{sec:comp-total-part}

The phase boundary for the W.I.D.B.G. model with static interactions
is peculiarly insensitive to the calculation scheme.
This can be seen by consider the Hugenholtz-Pines relation at the
transition.
In general, the anomalous self energy vanishes at the transition, so
$\mu = A(0,0)$.
Since $\rho_0=0$, the total density is $\rho_1$, which may be found
from the fluctuation Green's function, $\rho=\sum_{\omega,k}
\mathcal{G}(\omega,k)$,
\begin{equation}
  \label{eq:48}
  \mathcal{G}(\omega,k) = \left[
    i\omega + \epsilon_k - A(0,0) + A(\omega,k)
  \right]^{-1},
\end{equation}
where $\mu=A(0,0)$ has been used.
If the self energy is static, $A(\omega,k)=A(0,0)$, then at the transition
the quasi particles are exactly free.  
Any approximation scheme which gives $B(0,0)=0$ when $\rho_0=0$
will then reproduce this result.
For this reason, partial derivatives will give correct calculations of
the phase boundary for a dilute Bose gas, but this does not remain
true for dynamic self energies.

For a boson-fermion model, such as polaritons, because of the dynamic
self energy, total and partial derivatives will give different
answers.
For the three-dimensional case studied by Ohashi and Griffin, this
will lead to critical temperatures differing by a numerical factor,
but in two dimensions using partial derivatives gives divergent
answers.
Were one to use partial derivatives, the density calculated from the
condensed and non-condensed phases would agree at the critical chemical
potential.
However, for the total derivative, the density calculated at the new
critical potential need not agree with that calculated from the
non-condensed phase.
A difference between these results reflects the fact that both are
approximations of the phase boundary, and is indicative of the
Ginzburg criterion.

\begin{widetext}

\subsection{Total density for condensed polaritons}
\label{sec:total-dens-cond}

From the effective action, eq.~(\ref{eq:12}), the free energy per unit
area, including quadratic fluctuations may be written as:
\begin{eqnarray}
  \label{eq:49}
  \frac{F}{A}
  &=&
  \frac{S[\psi_0]}{A} 
  +
  \frac{1}{\beta}\int_0^{\infty}
  \frac{d^2 k}{(2\pi)^2} 
  \sum_{\omega} 
  \ln \left( 
    \left| i\omega + \hbar\omega_k + K_1(\omega) \right|^2 - 
    \left|K_2(\omega)\right|^2
  \right)
  \\
  \label{eq:50}
  &=& 
  \left\{
    \hbar \omtil_k \frac{|\psi_0|^2}{A} - \frac{\mu n}{2} - \frac{n}{\beta} 
    \ln\left[\cosh \left(\beta E \right) \right]
  \right\}
  +
  \left\{
    \frac{1}{\beta}
    \int_0^{\infty} 
    \frac{d^2 k}{(2\pi)^2} 
    \ln\left[ 1-e^{-\beta \hbar \omtil_k} \right]
  \right\}
  \nonumber\\
  &+& 
  \frac{1}{\beta}\int_0^{K_{\mathrm{m}}}
  \frac{d^2 k}{(2\pi)^2} \left\{
    \ln\left[
      \frac{
        \sinh(\beta \xi_1 /2)
        \sinh(\beta \xi_2 /2)
      }{
        \sinh(\beta E)
        \sinh(\beta \hbar \omtil_k/2)
      } \right]
    + 
    \frac{1}{2} \ln \left[
      1 - \frac{
        \alpha 4E^2 
      }{
        2 \left(
          \hbar \omtil_k E^2 - \hbar \omtil_0 \epstil^2 
        \right) } 
    \right]
  \right\}.
\end{eqnarray}
\end{widetext}

Here $E$, $\alpha$ and $\xi_{1,2}$ are defined as in
section~\ref{sec:greens-funct-fluct}.
The first term in braces is $S[\psi_0]$, the second and third
together are the fluctuation corrections.
As discussed in section~\ref{sec:high-energy-prop}, the interaction
has been cut off at a scale $K_{\mathrm{m}}$, so for
$k>K_{\mathrm{m}}$, the action is that of a free gas of photons,
{\it i.e.}\ the second term in braces.
In the third term, there are contributions both due to the Matsubara sum of
eq.~(\ref{eq:16}), and due to the $\delta_{\omega}$ terms in
eq.~(\ref{eq:12}), as discussed further in the
appendix~\ref{sec:mats-summ-with}.

\begin{widetext}
The total density is then given by:
\begin{equation}
  \label{eq:52}
  \rho
  =
    \frac{\left|\psi_0\right|^2}{A} +
    \frac{n}{2} \left[
      1 -  \frac{\epstil}{E} \tanh(\beta E)
    \right]
    +
    \int_0^{K_{\mathrm{m}}}\!\!\!
    \frac{d^2 k}{(2\pi)^2} 
    \left\{
      f[\xi_1] + f[\xi_2] - f[2E] +\frac{1}{2} + g(k)
    \right\}
  +
    \int_{K_{\mathrm{m}}}^{\infty}\!
    \frac{d^2 k}{(2\pi)^2} 
    n_{\mathrm{B}}(\hbar \omtil_k),
  \end{equation}
\end{widetext}
where
\begin{eqnarray}
  \label{eq:15}
    f[x]&=&
    \left( 
      n_{\mathrm{B}}(x) + \frac{1}{2} 
    \right) 
    \left(
      -\frac{d x}{d\mu}
    \right)
    ,
    \nonumber\\
    g(k) &=& -\frac{1}{2\beta} \frac{1}{(1-C)} \frac{d C}{d\mu}
    , 
    \nonumber\\
    C&=&\frac{
      \beta \sech^2(\beta E) g^2 n
    }{
      2 \left(
        \hbar \omtil_k E^2 - \hbar \omtil_0 \epstil^2 
      \right) } 
    \frac{ g^2 |\psi_0|^2}{A}. 
    \nonumber  
\end{eqnarray}

In going from eq.~(\ref{eq:50}) to eq.~(\ref{eq:52}) the two integrals
have been re-arranged, the second now describing only the free, high
energy photons. 
Again, the last term, $g(k)$, arises due to the $\delta_{\omega}$
terms.

The derivatives of polariton energies that arise in calculating the
density may be given in terms of the expressions $A(k)$, $B(k)$ as
defined in eq.~(\ref{eq:17}):
\begin{eqnarray}
  \label{eq:53}
  \frac{2\xi_{1,2}}{d\mu}
  &=&
  \frac{1}{4 \xi_{1,2}} 
  \left[
    \frac{d A(k)}{d \mu}
    \pm
    \frac{1}{\sqrt{A(k)^2 - B(k)}}
    \times
  \right.
  \nonumber\\
  && \qquad
  \times
  \left.
    \left(
    2 A(k) \frac{d A(k)}{d \mu}
    -
    \frac{d B(k)}{d \mu}
    \right)
  \right],
  \\
  \label{eq:54}
  \frac{d A(k)}{d \mu}
  &=&
  8 E \frac{d E}{d \mu}
  -
  2 \hbar \omega_k
  -
  4\epstil
  -
  2 \hbar \omtil_0,
  \\
  \label{eq:55}
  \frac{d B(k)}{d \mu}
  &=&
  16 \frac{\hbar^2 k^2}{2m} \left(
    2 E \frac{d E}{d \mu} \omtil_k
    -
    E^2
    + 
    \epstil \hbar \omtil_0
    + 
    \epstil^2
  \right).
\end{eqnarray}
To find $d E/d \mu$ in the presence of a condensate, one can
differentiate the gap equation, eq.~(\ref{eq:7}), giving
\begin{equation}
  \label{eq:56}
  -1=\frac{dE}{d\mu} \frac{g^2 n}{2 E^2}\left(
    \beta E \mathrm{sech}^2(\beta E)
    -
    \tanh(\beta E)
  \right).
\end{equation}

\subsection{Two dimensions, Superfluid response}
\label{sec:two-dimens-superfl}

Having found an expression for the total density including
fluctuations, it is necessary to consider how fluctuations
change the critical chemical potential.
As discussed in the introduction to this section, in two dimensions
this requires consideration of the Kosterlitz-Thouless phase
transition.
The phase boundary is approximated from the condensed state by the
chemical potential at which the superfluid response vanishes.
We therefore must calculate the normal and superfluid response in the
presence of a condensate.

\subsubsection{Calculating normal response density}
\label{sec:calc-norm-resp}

Following the standard procedure%
\cite{griffin94:_excit_bose_liquid,pitaevskii03:_BEC},
we consider the current,
\begin{equation}
  \label{eq:57}
  \vec{J}(\vec{q},0)=\sum_{\vec{k},\omega} \frac{\hbar \vec{k} }{m}
  \psi^{\dagger}_{\vec{k}-\vec{q}/2}
  \psi^{}_{\vec{k}+\vec{q}/2}.
\end{equation}
For a perturbation $\delta H = \hbar \delta \vec{l}.\vec{J}$, the linear response 
may be written
$\langle J_i(q,0) \rangle = \chi_{ij}(q)\delta l_j(q)$.
By symmetry, the most general response function is,
\begin{equation}
  \label{eq:58}
  \chi_{ij}(\vec{q}) =
  \chi_{\mathrm{L}}(\vec{q})
  \frac{q_i q_j}{q^2}
  +
  \chi_{\mathrm{T}}(\vec{q})
  \left(\delta_{ij} -   \frac{q_i q_j}{q^2} \right).
\end{equation}
By gauge symmetry, eq.~(\ref{eq:57}) is a conserved current.
It follows, via a Ward identity,  that 
\begin{math}
  q_i\chi_{ij}(q) = q_j \mathrm{Tr} (\mathcal{G}(k)),
\end{math}
so the total density is 
\begin{math}
  \rho = m \chi_{\mathrm{L}}(\vec{q}\rightarrow 0) /A
\end{math}.
In contrast, the transverse response depends on only the density of normal
particles, 
\begin{math}
  \rho_{\mathrm{normal}} =m\chi_{\mathrm{T}}(\vec{q}\rightarrow 0) /A 
\end{math}.

In order to calculate the total density correctly, it is necessary
to introduce vertex corrections, {\it i.e.}\
\begin{equation}
  \label{eq:59}
  \chi_{ij}(\vec{q}\rightarrow 0) = 
  \mathrm{Tr} \left(
    \Gamma_i(\vec{k},\vec{k}) \mathcal{G}(\vec{k})
    \gamma_j(\vec{k},\vec{k}) \mathcal{G}(\vec{k})
  \right),
\end{equation}
where $\gamma_i(\vec{p},\vec{q}) = \sigma_3(p_i+q_i)/2m$, and $\Gamma_i(p,q)$ 
is chosen to satisfy the Ward identity.
However, the diagrams required to satisfy the Ward identity (see
ref.~\onlinecite{griffin94:_excit_bose_liquid}) take the form,
\begin{equation}
  \label{eq:60}
  \Gamma_i(\vec{p}, \vec{q}) = 
  \gamma_i(\vec{p}, \vec{q}) + (p_i-q_i) f(p,q).
\end{equation}
This form of the vertex correction means only the longitudinal
response is affected.
Therefore, the standard procedure is to calculate the total density
directly from the free energy, as in
section~\ref{sec:total-dens-cond}, and the normal density by linear
response.

To one loop order, and neglecting vertex corrections, the response function
is given by,
\begin{eqnarray}
  \label{eq:61}
  \chi_{ij}(\vec{q\rightarrow0}) = \frac{1}{\beta}
  \sum_{\vec{k},\omega} \frac{\hbar^2 k_i k_j}{m^2}
  \mathrm{Tr} \left(
    \mathcal{G}(\vec{k}) \sigma_3
    \mathcal{G}(\vec{k}) \sigma_3
  \right),
\end{eqnarray}
and so, taking the continuum limit, the normal density is given by
\begin{equation}
  \label{eq:62}
  \rho_{\mathrm{n}} = \int_0^{\infty} \frac{d^2 k}{(2\pi)^2}
  \frac{\hbar^2 k^2}{2m}
  \frac{1}{\beta} \sum_{\omega}
  \mathrm{Tr} \left(
    \mathcal{G}(\vec{k}) \sigma_3
    \mathcal{G}(\vec{k}) \sigma_3
  \right).
\end{equation}

\begin{widetext}
  For the polariton system, the trace can be evaluated to give
  \begin{eqnarray}
    \label{eq:63}
    \rho_{\mathrm{n}} 
    &=&
    \int_0^{\infty} \frac{d^2 k}{(2\pi)^2}
    \frac{\hbar^2 k^2}{2m}
    \frac{1}{\beta}\sum_{\omega}
    \frac{
      \left(i\omega + \hbar \omtil_k + K_1(\omega)\right)^2 +
      \left(-i\omega + \hbar \omtil_k + K_1^{\ast}(\omega)\right)^2 -
      2 \left|K_2(\omega)\right|^2
    }{\left(
        \left|i\omega + \hbar \omtil_k + K_1(\omega)\right|^2 - 
        \left|K_2(\omega)\right|^2
      \right)^2
    }
    \\
    \label{eq:64}
    &=&
    \int_0^{\infty} \frac{d^2 k}{(2\pi)^2}
    \frac{\hbar^2 k^2}{2m}
    \frac{1}{\beta}
    \left \{
    \sum_{\omega}
    \frac{\left[
        2\tilde{\omega_0}(i\tilde{\epsilon}\omega - E^2 - \tilde{\epsilon}^2)
        +(i\omega + \tilde{\omega_k})(\omega^2+4E^2)
      \right]^2 -
      \left[2 \tilde{\omega_0}
        (E^2 -\tilde{\epsilon}^2)
      \right]^2
    }
    {(\omega^2+\xi_1^2)^2(\omega^2+\xi_2^2)^2}
    + 
    \mathrm{C_0}(k)
    \right\}.
  \end{eqnarray}    
  Again, in evaluating the Matsubara sum, one must consider the
  $\delta_{\omega}$ terms. 
  The term $\mathrm{C_0}(k)$ is the
  difference between the true term at $\omega=0$, and the analytic
  continuation appearing in the Matsubara sum in eq.~(\ref{eq:64}),
  and is given by:
  \begin{equation}
    \label{eq:88}
    \mathrm{C_0}(k)
    =
    2 \alpha 
    \times
    \left[
      \left(
        \frac{\hbar^2 k^2}{2m}
      \right)
      \left(
        \frac{\hbar^2 k^2}{2m}
        + 
        \frac{g^2 |\psi_0|^2}{A}
        \frac{\hbar \omtil_0}{E^2}
      \right)
      \left(
        \frac{\hbar^2 k^2}{2m}
        + 
        \frac{g^2 |\psi_0|^2}{A}
        \frac{\hbar \omtil_0}{E^2}
        - 
        2\alpha
      \right)
    \right]^{-1},
  \end{equation}
  with $\alpha$ as defined in eq.~(\ref{eq:93}).
\end{widetext}


\subsubsection{Total photon density}
\label{sec:total-photon-density}

For the polariton system there is an added complication.
Equation~(\ref{eq:64}) gives the density of normal photons, but
equation~(\ref{eq:52}) is the total excitation density (including
excitons).
It is therefore necessary to calculate the total photon density.

This can be done by considering separate chemical potentials for
photons and excitons, which are set equal at the end of the
calculation.
This means making the change,
\begin{equation}
  \label{eq:65}
  \mu N 
  \rightarrow
  \mu_{\mathrm{ex.}}
  \sum_{j=1}^{j=nA} \left(
    S_j^z+ \frac{1}{2}
  \right)
  + \mu_{\mathrm{phot.}}
  \sum_{k} \psi^{\dagger}_k \psi^{}_k,
\end{equation}
in the action.
The photon density is then total derivative w.r.t.\ the photon
chemical potential, $\mu_{\mathrm{phot.}}$.

This density is given by equation~(\ref{eq:52}) with two changes.
Firstly,  the mean-field exciton density,
\begin{equation}
  \label{eq:66}
    \frac{n}{2} \left[
      1 -  \frac{\epstil}{E} \tanh(\beta E)
    \right],
\end{equation}
should be removed.
Secondly, in $f[x]$,$g(k)$ derivatives should be taken w.r.t\ 
$\mu_{\mathrm{phot.}}$. 
This means replacing equations~(\ref{eq:54}) and (\ref{eq:55}) by:
\begin{eqnarray}
  \label{eq:67}
    \frac{d A(k)}{d \mu_{\mathrm{phot.}}}
  &=&
  8 E \frac{d E}{d \mu}
  -
  2 \hbar \omega_k
  -
  4\epstil,
  \\
  \label{eq:68}
  \frac{d B(k)}{d \mu_{\mathrm{phot.}}}
  &=&
  16 \frac{\hbar^2 k^2}{2m} \left(
    2 E \frac{d E}{d \mu} \omtil_k
    -
    E^2
    + 
    \epstil^2
  \right).
\end{eqnarray}
This makes use of the fact that $d E/d \mu_{\mathrm{phot.}}=d E/ d
\mu$, as can be seen from the gap equation.

\section{Phase boundary including fluctuations}
\label{sec:new-phase-boundary}

\begin{figure}[htbp]
  \centering
  \includegraphics[width=3.4in]{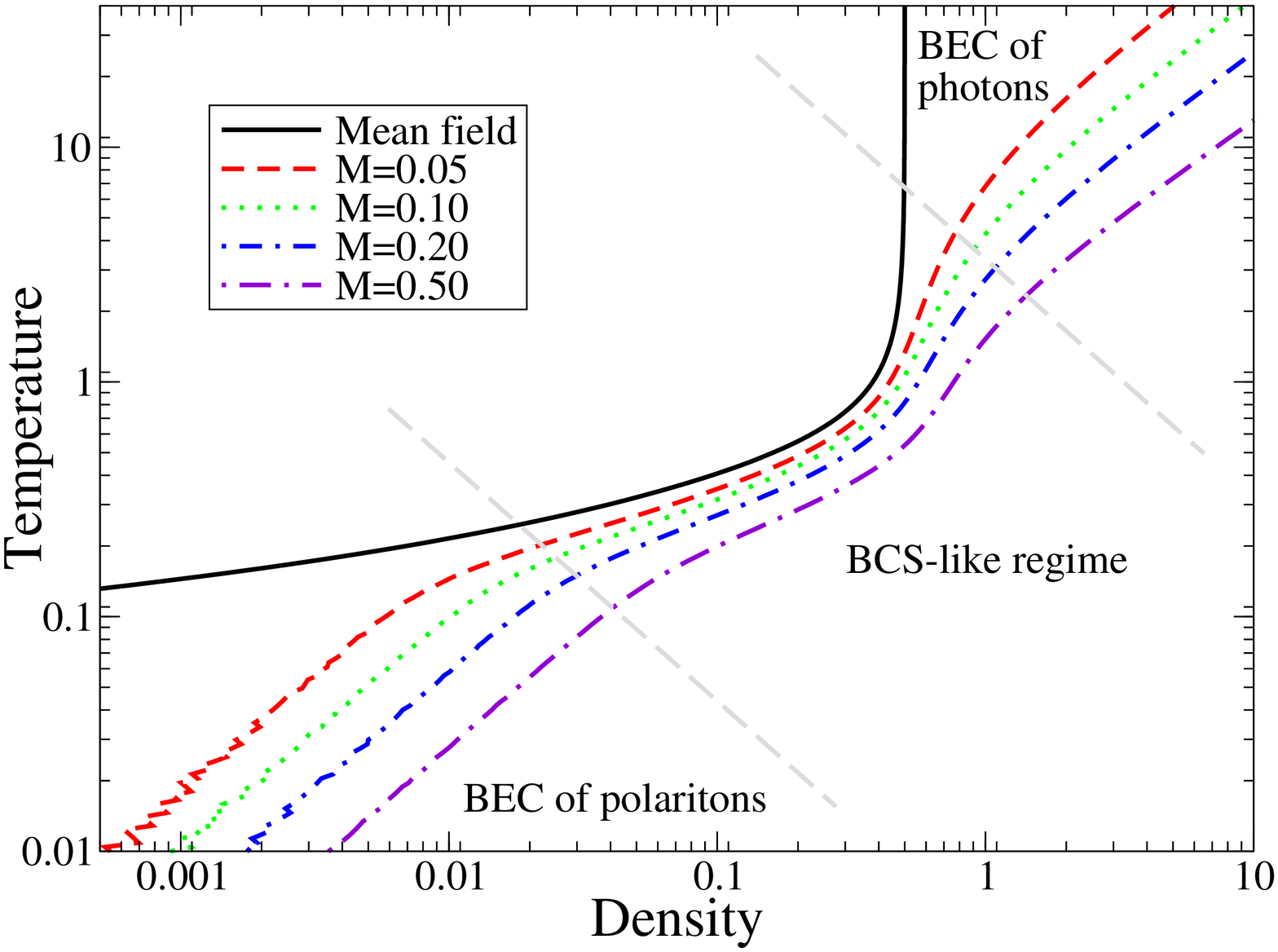}
  \includegraphics[width=3.4in]{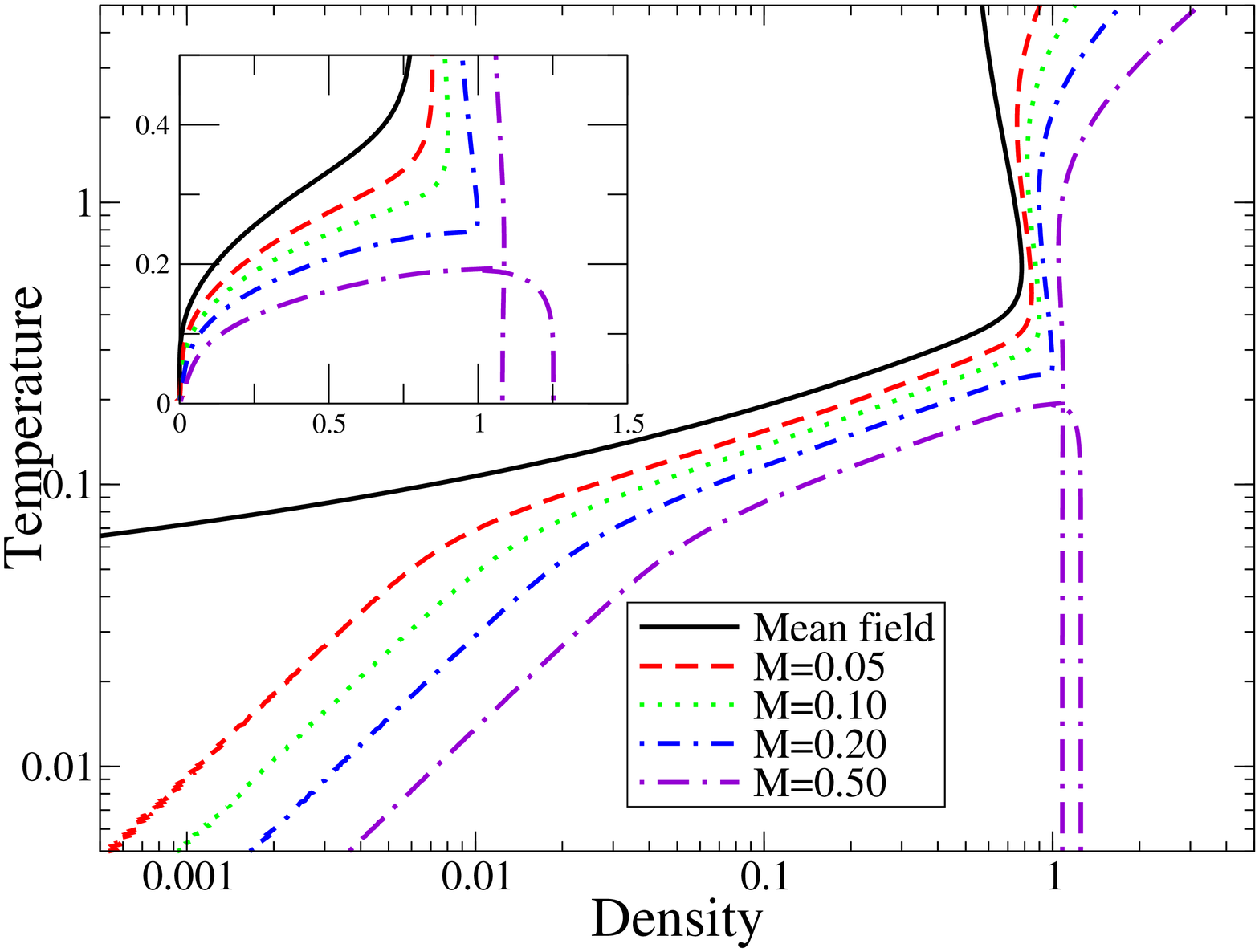}
  \includegraphics[width=3.4in]{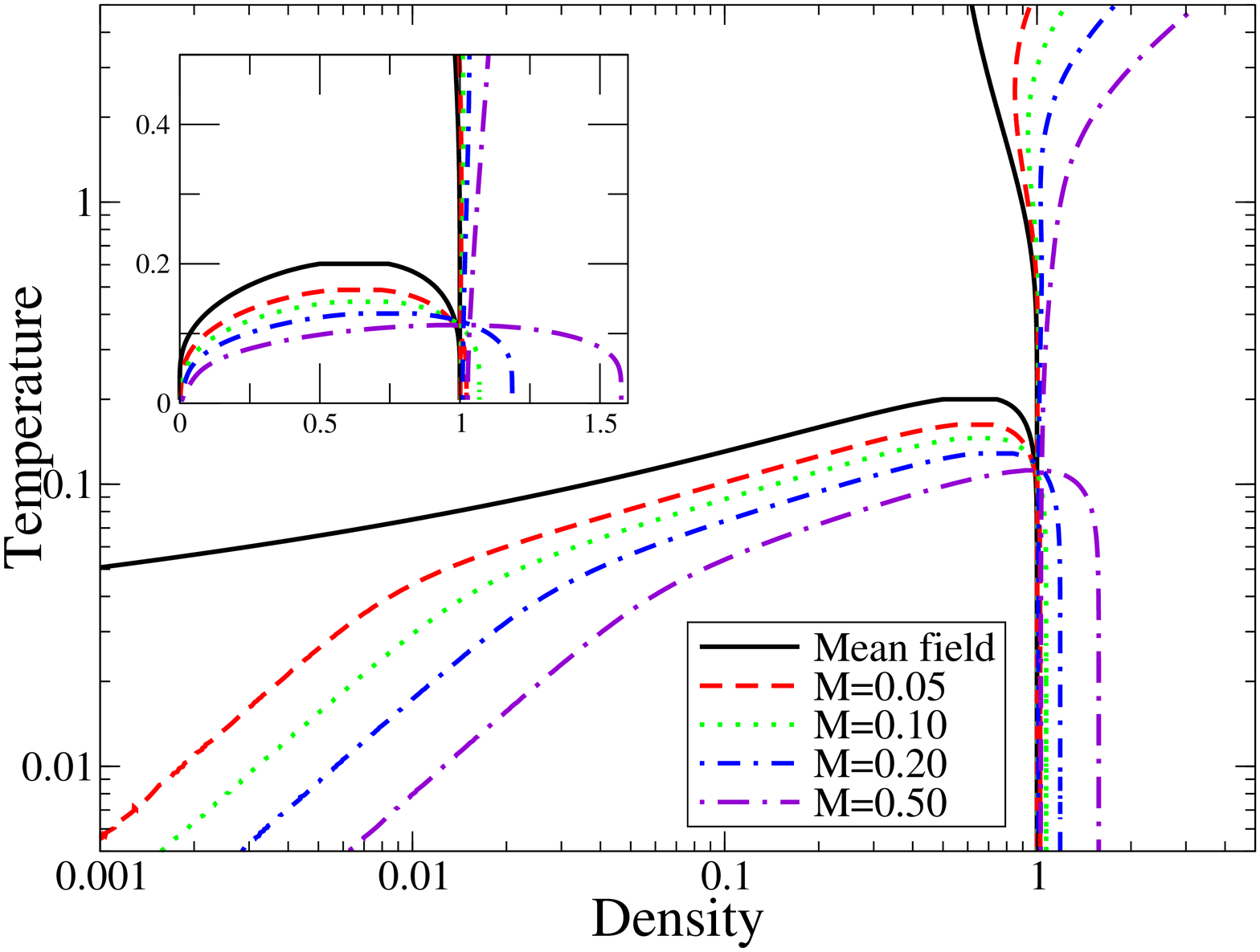}
  \caption{Mean-field phase boundary, and phase boundaries including
    fluctuation correction for four values of photon mass, on a
    logarithmic scale (Insets are on a linear scale).
    The top panel is the resonant case, $\Delta=0$, adapted from
    ref.~\onlinecite{keeling04:_polar} and shown for comparison.
    In the lower two panels the exciton is detuned below the photon
    band by $\Delta=1.5g\sqrt{n}$ and $\Delta=2.5g\sqrt{n}$ respectively.
    Temperature plotted in units of $g\sqrt{n}$ and density in units of $n$.
  }
  \label{fig:phasediag}
\end{figure}

Combining the results of section~\ref{sec:fluct-corr-dens}, the phase
boundary is found by plotting the total density (eq.~(\ref{eq:52})) at
the value of chemical potential where the normal photon density
(eq.~(\ref{eq:64})) matches the total photon density (discussed in
section~\ref{sec:total-photon-density}).
The phase boundaries found in this way are plotted in
figure~\ref{fig:phasediag}.
The form of the phase boundary can be explained by considering how, at
finite temperatures, the occupation of excited states of the system
depletes the condensate. 
Which excited states are relevant changes with density.

\subsection{Resonant case}
\label{sec:resonant-case}

When condensed, the lowest energy mode is the phase mode, described
by eq.~(\ref{eq:22}).  At low density, this has a shallow slope,
and consequently a large density of states.
Such excitations are described in a model of point bosons.
The phase boundary can therefore be estimated from the degeneracy
temperature of a gas of polaritons, of mass $2m$, where $m$ is the
bare photon mass:
\begin{equation}
  \label{eq:69}
  T_{deg} 
  = 
  \frac{2\pi \hbar^2}{2 (2m)} \rho
  =
  g\sqrt{n} \frac{\pi}{2 m^{\ast}} \frac{\rho}{n}.
\end{equation}

As the density increases, the phase mode becomes steeper, and so
has a smaller density of states.  The relevant excitations are then
single particle excitations across the gap.
Such excitations are accounted for in the mean-field theory.
Combining equations (\ref{eq:7}) and (\ref{eq:95}) gives the result:
\begin{equation}
  \label{eq:70}
    T_c =  
  g\sqrt{n} \frac{\sqrt{1-2\rho/n}}{2 \tanh^{-1}(1-2\rho/n) } 
  \approx
  \frac{g\sqrt{n}}{-\ln (\rho/n)}.
\end{equation}

As seen in figure~\ref{fig:phasediag}, the mean-field boundary is
effectively constant on the scale of the boundary for BEC of point
bosons, and so the crossover to mean-field always occurs near
$T\approx g\sqrt{n}$, the Rabi splitting.
The density at which this crossover occurs depends on the
photon mass.  
Comparing equations (\ref{eq:69}) and (\ref{eq:70}), this crossover
occurs at a density $\rho_{\mathrm{crossover}} \approx n m^{\ast}$.

In terms of the measurable Rabi splitting, $g\sqrt{n}$ and polariton
mass $m$, this gives the density:
\begin{equation}
  \label{eq:71}
  \rho_{\mathrm{crossover}} = 
  \frac{m g \sqrt{n}}{\hbar^2} 
\end{equation}
For the structures studied by Yamamoto
{\it et  al.}\cite{yamamoto02:_condensation,deng03:_polar,weihs04:_polar}, %
$g\sqrt{n}\approx 7 \mathrm{meV}$ and $m\approx 10^{-5} m_{\mathrm{electron}}$.
These values give a crossover density of
$\rho_{\mathrm{crossover}}=2.6\times10^{8} \mathrm{cm}^{-2}$.
This is both much less than the estimates of experimentally
achieved density, $n \sim 10^{11} \mathrm{cm}^{-2}$ per pulse, and
also much less than the Mott density in this structure,
$n_{\mathrm{Mott}}\approx 3.6 \times 10^{13} \mathrm{cm}^{-2}$.
For the structures studied by Dang {\it et al.}%
\cite{dang98:_stimul,richard04:_angle_cdte_ii_vi}
$g\sqrt{n} \approx 13\mathrm{meV}$, and $m\approx 3\times10^{-5}
m_{\mathrm{electron}}$, so crossover densities are again of the same
order, $\rho_{\mathrm{crossover}}=5\times10^{8} \mathrm{cm}^{-2}$.
Dang {\it et al.}\ also presents results for the detuned
case\cite{richard04:_angle_cdte_ii_vi}, discussed below, with a range
dimensionless detunings $0.5 > \Delta^{\ast} > - 0.7$.

Equation~(\ref{eq:71}) describes the crossover in terms of properties
measurable for a given microcavity. 
However, to understand what fundamental lengthscales control this
crossover density, it is necessary to write the coupling strength and
polariton mass in terms of the dimensions of the cavity and properties
of the excitons.
Using the expressions in section~\ref{sec:model-parameters} for photon
mass and coupling strength $g$, this gives the crossover density as:
\begin{equation}
  \label{eq:80}
  \rho_{\mathrm{crossover}} = 
  4\pi^2 \sqrt{\frac{e^2}{4\pi \varepsilon_0 \hbar c}}
  \frac{1}{\varepsilon_r^{1/4}}
  \frac{d_{ab} \sqrt{n}}{w^2}.
\end{equation}

The crossover density is therefore controlled by two parameters: The
width of the cavity, $w$, and the ratio of electron-hole separation to
average two-level system separation,
$d_{ab}\sqrt{n}=d_{ab}/r_{\mathrm{separation}}$.
If the average two-level system separation ($r_{\mathrm{separation}}$)
is less than the electron-hole separation ($d_{ab}$), then our model
of localised two-level systems will break down.
Therefore, within our model the largest possible crossover density
scale is $1/w^2$.
This lengthscale occurs because the cavity size controls the
wavelength of the lowest radiation mode.
Crossover to a BCS-like mean-field regime occurs when the density
approaches a scale set by the wavelength of light, rather than one set
by the exciton Bohr radius.
Therefore, in general this crossover density is much less than the
Mott density.

\begin{figure}[htbp]
  \centering
  \vspace{2ex}
  \includegraphics[width=1.1in]{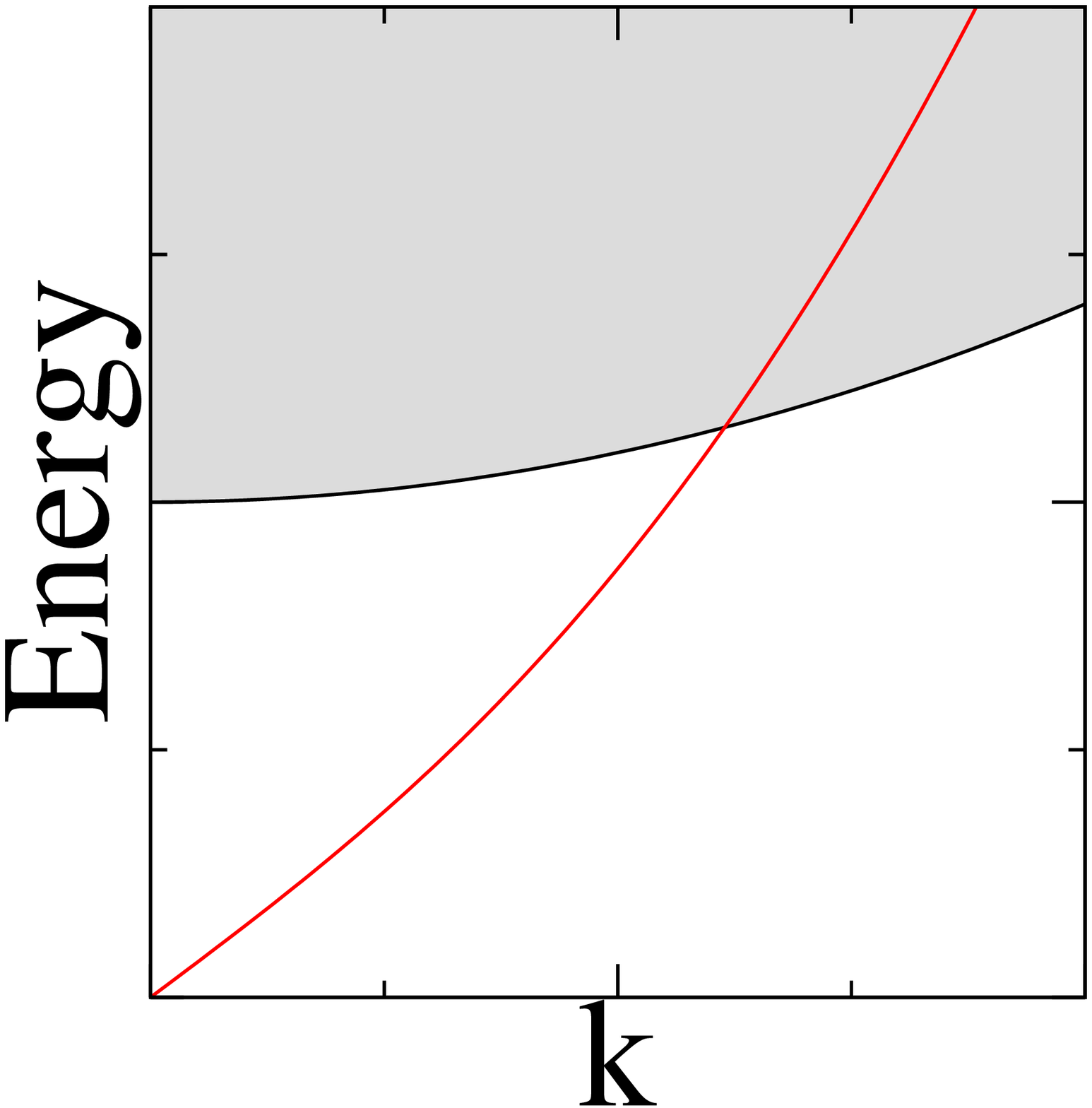}
  \includegraphics[width=1.1in]{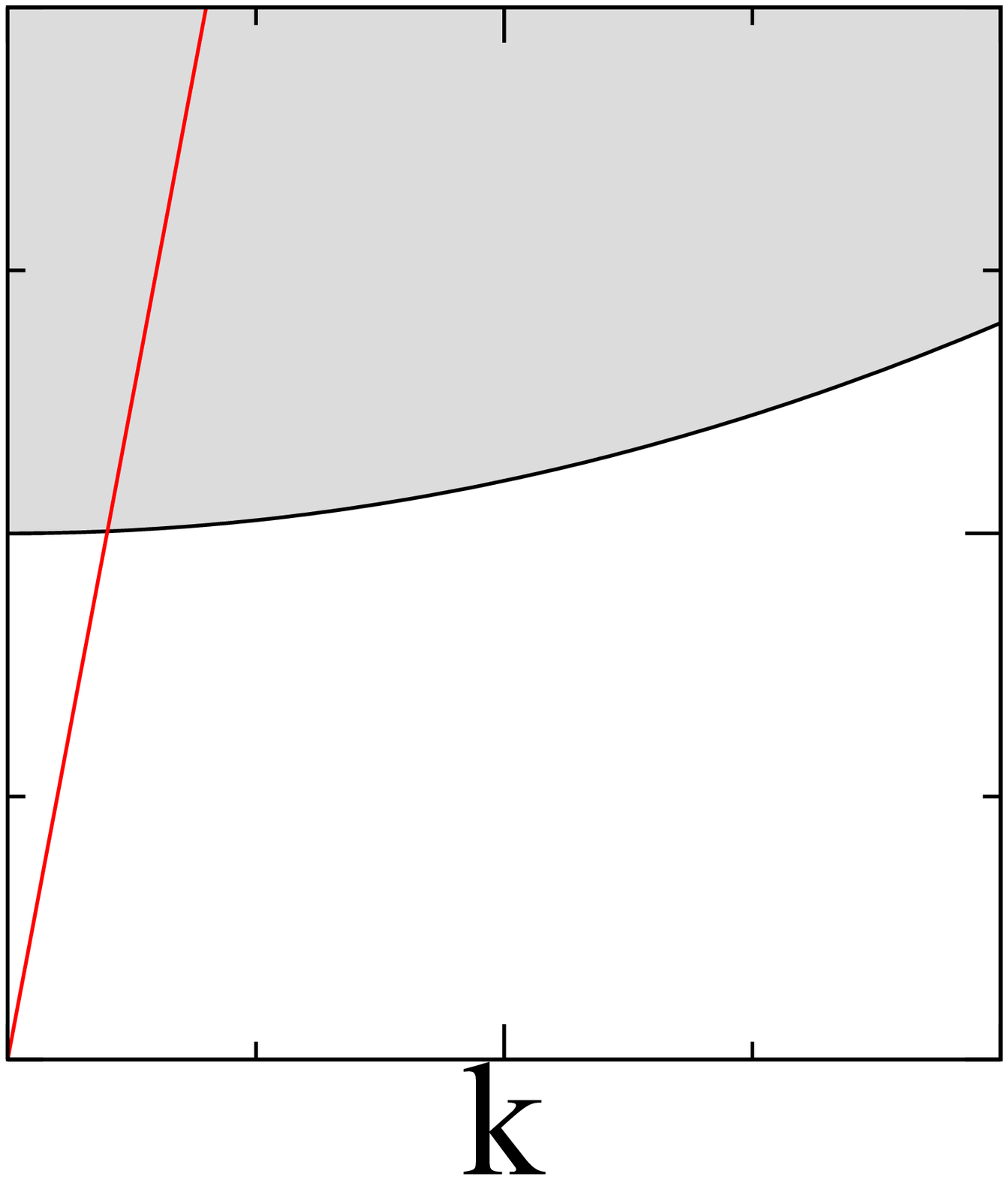}
  \includegraphics[width=1.1in]{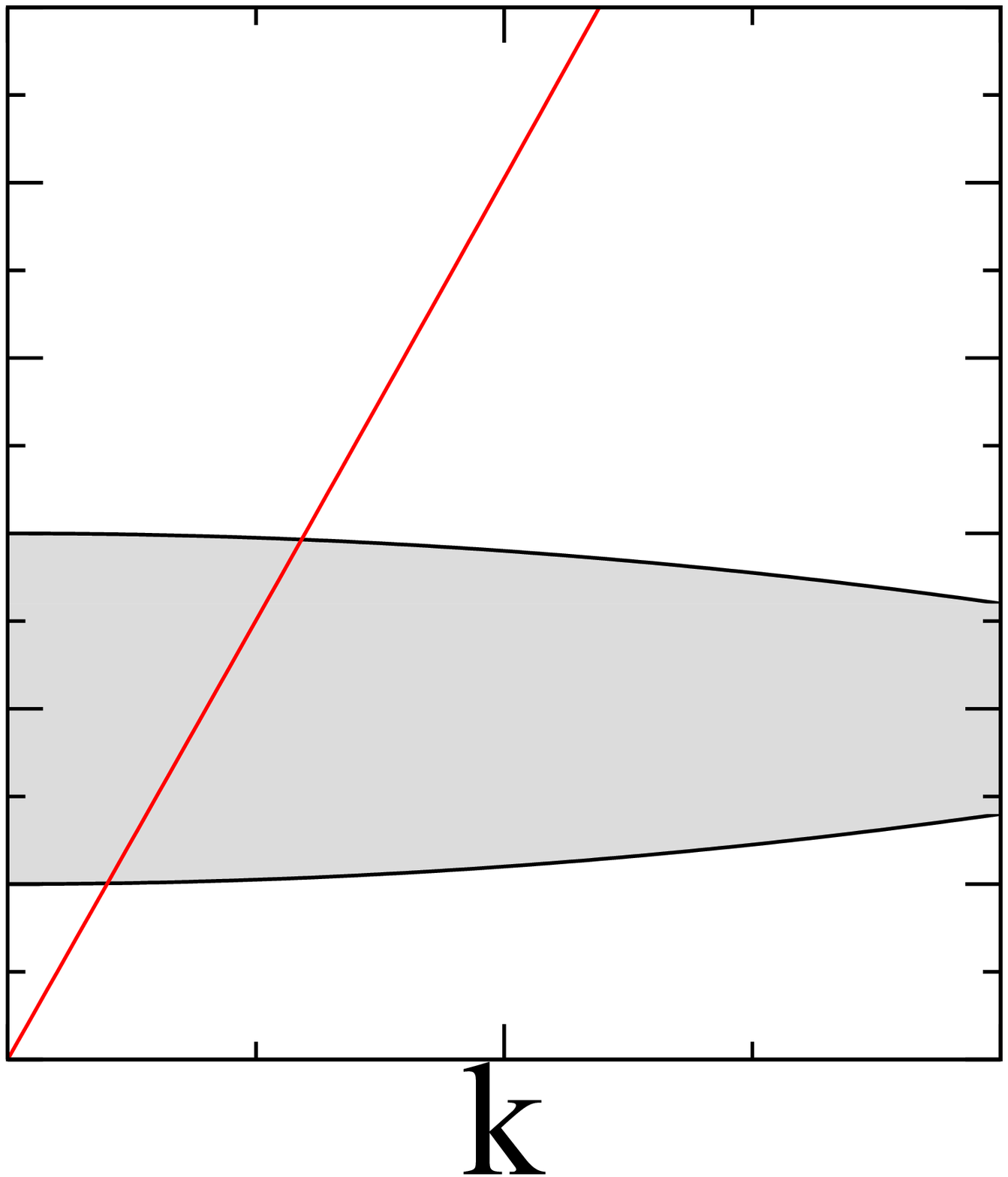}
  \caption{A schematic picture of how the relevant excitations change 
    between the polariton BEC, the BCS-like mean-field and the photon
    BEC regimes.
    In the low density limit, a shallow sound mode exists.
    At higher densities, this becomes steeper, and the relevant
    excitations are gapped single particle excitations.
    At yet higher densities, these modes are saturated, and the high
    $k$ photon modes become relevant.
  }
  \label{fig:schematic}
\end{figure}

At yet higher densities, the single particle excitations are saturated,
and so the condensate becomes photon dominated.
In this regime, the transition temperature is that for a gas of
massive photons.
If the photon mass is large ($m^{\ast} > 1$), a mean-field regime
never exists, instead the phase boundary changes directly from 
polariton condensation to a photon condensation.
However, since for experimental parameters the dimensionless mass is
only of the order of $10^{-3}$, a mean-field regime will exist.
These various crossovers are illustrated schematically in
figure~\ref{fig:schematic}.

\subsection{Effects of detuning}
\label{sec:effects-detuning}

If the excitons are detuned below the photon (positive detuning),
it becomes possible for the system to reach half filling while
remaining uncondensed.
For positive detunings greater than $2g \sqrt{n}$ the mean-field phase
boundary becomes re-entrant, as shown in the bottom panel of
figure~\ref{fig:phasediag}.
For smaller but still positive detunings, the mean-field boundary has
a maximum critical density at a finite temperature, but no maximum of
critical temperature.
The opposite case, of excitons detuned above photons, shows no
interesting features; the system will always condensed before
half filling.

This multi-valued phase boundary is discussed in
ref.~\onlinecite{eastham01:_bose}, and can be explained in terms of
phase locking of precessing spins\cite{eastham03:_phase}, either about
spin down (low density) or spin up (high density) states.
Above inversion, increasing the density reduces the extent to 
which a spin may precess.
For very large detunings, at low temperatures, the phase diagram
therefore becomes symmetric about half filling.

When $\Delta \geq 2 g\sqrt{n}$, the re-entrance leads to a point at
zero temperature where two second order phase boundaries meet.
Including fluctuations, as shown in the bottom panel of
figure~\ref{fig:phasediag}, these phase boundaries no longer meet, but
instead there is a region where two different condensed solutions
coexist.
In this region, there are two minima of the free energy, so we expect
there will be a first-order phase boundary between them.
Although this boundary could be calculated by comparing the free
energies including fluctuations, its form may be altered significantly
by higher order corrections.

\begin{figure}[htbp]
  \centering
  \includegraphics[width=3.4in]{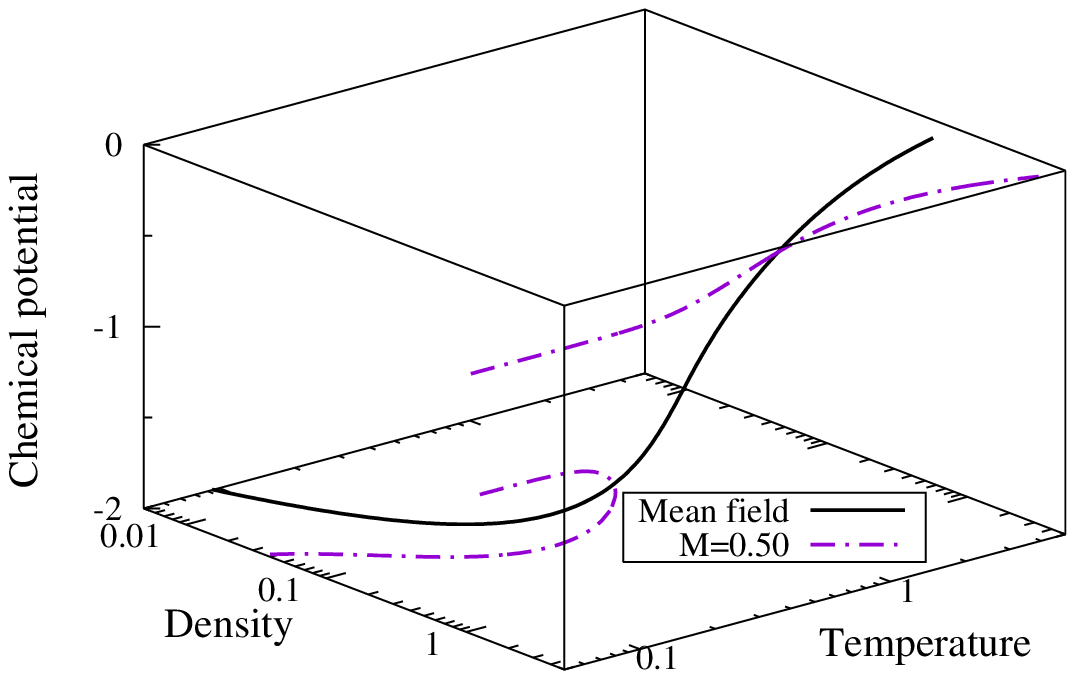}
  \includegraphics[width=3.4in]{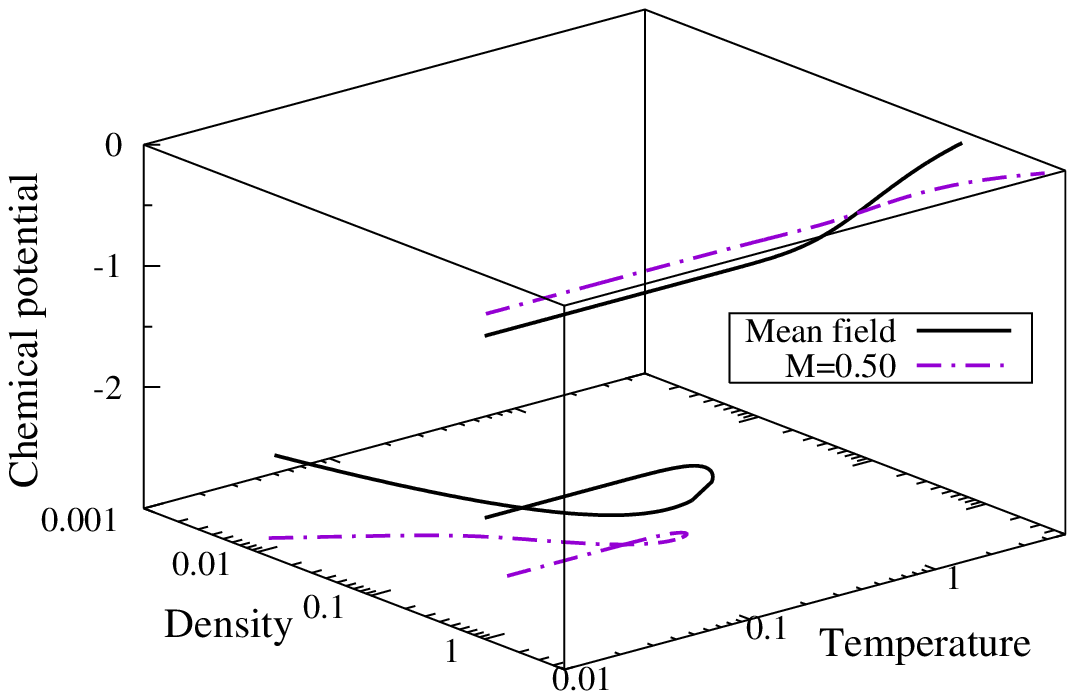}
  \caption{Chemical potential vs density and temperature
    at the phase boundary, for the mean-field phase boundary, and
    $m^{\ast}=0.50$ fluctuation corrections.
    Plotted for detunings of $\Delta=1.5g\sqrt{n}$ and
    $\Delta=2.5g\sqrt{n}$, with all other parameters as in
    figure~\ref{fig:phasediag}.
    Temperature and chemical potential plotted in units of
    $g\sqrt{n}$, and density in units of $n$.  }
  \label{fig:mu-rho-T}
\end{figure}

In the mean-field theory, at zero temperature, the chemical potential
jumps discontinuously at the point where the two phase boundaries
meet.
This can be understood by the chemical potential locking to the lower
polariton for the lower density transitions, and to the upper polariton
at higher densities.
This can be seen in the lower panel of figure~\ref{fig:mu-rho-T},
which plots the value of the chemical potential at the phase boundary.
Including fluctuations, the jump in chemical potential has a similar
form, and is somewhat larger.
In the region of coexistence discussed above, the two minima of free
energy have different chemical potentials, so at the first-order
transition, the chemical potential will jump.

At smaller detunings, as shown in the central panel of
figure~\ref{fig:phasediag}, although the mean-field phase boundary is
single valued, adding fluctuations can reproduce the same coexistence
regions.
For this to occur, the photon mass must be large --- {\it i.e.}\ there
must be a significant density of states for fluctuations.
As shown in the upper panel of figure~\ref{fig:mu-rho-T}, this
coexistence is also characterised by two minima of the free energy,
with different chemical potentials, and so is also expected to become
a first-order transition in the same manner.

To explain how fluctuations lead to the introduction of multiple phase
boundaries at a single temperature, it is necessary to consider the
upper branch of excitations, $\xi_2(k)$.
With positive detuning, the energy of this mode (w.r.t.\ chemical
potential) can continue to fall as the chemical potential increases in
the condensed state.
This has two effects, it makes the sound velocity larger (as can be
seen from eq.~(\ref{eq:22})), and increases the population of this
``pair-breaking'' upper mode.
The combination of these effects is responsible for the creation of
the coexistence region by fluctuations.

\subsection{Effects of inhomogeneous broadening}
\label{sec:effects-inhom-broad}

It is interesting to consider how a small inhomogeneous broadening
will modify the phase boundary.
Exact calculations with a continuum of exciton energies are
technically challenging and not particularly illuminating, so the
following presents a discussion of the main effects expected.
The following discussion is for a Gaussian distribution of energies,
centred at the bottom of the photon band, with a standard deviation
much less than $g\sqrt{n}$.

The most significant change to the boundary is due to the existence of
a low energy tail of excitons.
This means that, even at low densities, the chemical potential lies
within the exciton band, and a BCS-like form of $T_c$ will be
recovered.
Consider the density of states,
\begin{equation}
  \label{eq:73}
  \nu_{s}(\epsilon)=
  \frac{\exp\left(-\epsilon^2/2\sigma^2 \right)}
  {\sqrt{2\pi}\sigma}.
\end{equation}
For large negative chemical potentials, at low temperatures and
densities, the saddle point equation~(\ref{eq:7}) becomes
\begin{eqnarray}
  \label{eq:72}
  \frac{\hbar \omega_0}{g^2}
  =
  \frac{1}{g_{\mathrm{eff}}}
  &=&  
  \int_{-\infty}^{\infty}
  \frac{\tanh(\beta \epstil)}{2\epstil}
  \nu_{s}(\epsilon)
  d\epsilon,
  \nonumber
  \\
  &\approx&
  \nu_{s} \left( \frac{\mu}{2} \right)
  \left[
    1 + \ln\left(\frac{\Lambda}{T} \right)
  \right],
\end{eqnarray}
and the mean-field density~(\ref{eq:95}), using the asymptotic form of
the error function, is
\begin{equation}
  \label{eq:74}
  \rho_{\mathrm{M.F.}} 
  =
  \nu_{s}\left( \frac{\mu}{2} \right) \left(\frac{\sigma^2}{-\mu}\right).
\end{equation}

The cutoff, $\Lambda$ is approximately $2 \sigma^2/\mu$, but appears
only as a pre-exponential factor, and so the density dependence it
gives to $T_c$ will be neglected.
Thus, the mean-field transition temperature at low densities then becomes
\begin{equation}
  \label{eq:75}
  T_c = \Lambda \exp\left( - \frac{2\sigma^2}{g^2 \rho}\right).
\end{equation}

For low densities, this result is very different to the mean-field
theory without broadening, eq.~(\ref{eq:70}).
Whereas before the mean-field boundary was approximately constant, it
now drops rapidly to zero.
If one now considers how fluctuations will modify this boundary, it is
more helpful to consider the density as a function of temperature.
At low temperatures, fluctuations increase the density by a small
amount, $\Delta \rho \propto T$.
Without broadening, the mean-field critical density is approximately
$\rho \approx n e^{-g\sqrt{n}/T}$ and goes to $0$ faster than the
fluctuation corrections, $\Delta\rho$.
Therefore, as shown in figure~\ref{fig:phasediag}, at low temperatures
the fluctuation contribution controls the phase boundary.
With broadening the mean-field critical density at low temperature is
approximately $\rho \approx 2\sigma^2/g^2\ln(\Lambda/T)$, which goes
to zero more slowly than $\Delta \rho$.
Therefore, including fluctuations in this regime does not change the
form of the phase boundary drastically.
Hence at very low densities, the boundary is again well described by a
mean-field theory.

\subsection{Relation to alternate models and experimental systems}
\label{sec:relat-exper-syst}

Our model describes two-level systems with a finite total density of
states --- the density of states, per unit area, integrated over all
energies, is finite.
As will be explained below, this finite total density of states is
responsible for the re-entrant behaviour seen in the mean-field theory
for detunings $\Delta \ge 2 g\sqrt{n}$.
In the preceding sections, the finite total density of states has also been
implicated in explaining the existence of a photon dominated region at
high temperature, and the multi-valued phase boundary in the
presence of fluctuations.
This section discusses how these effects may change in alternate
models which do not have saturable two-level systems.
Although the multi-valued phase boundary is expected only to occur for
a finite band of two-level systems, the existence of a photon
dominated regime at high densities is more general.

Before discussing the more involved question of how changing the
density of states affects fluctuations, we first summarise its effect
on the mean-field
theory\cite{eastham00:_bose,eastham01:_bose,littlewood04:_model}.
Consider the highest possible density achievable in the normal state.
Since there exists a bosonic mode, the chemical potential cannot
exceed the energy of this mode if the system is to remain normal.
Therefore at zero temperature, only exciton modes below the
boson mode are relevant.
Regardless of whether the total density of states is finite, the
density of states below the bosonic mode will be finite.
However, at non-zero temperatures, exciton modes above the chemical
potential can be occupied by the tail of the Fermi distribution.
If there is a finite total density of states, there will be a maximum
density of excitons that can be occupied thermally.
This is what is referred to as ``saturation'' below.
Note that for a exciton band centred at the bottom of the photon band,
this maximum density of excitons is half the density of two-level
systems --- the system saturates at half filling.
If the total density of states is not finite, it is possible to
support any total density in the normal state by making the
temperature high enough.

This saturation is responsible for the multi-valued phase boundary
in the mean-field theory.
If the density is close to total inversion of the two-level systems,
all two-level systems must be in the up state.
This makes it hard for them to produce a macroscopic polarisation ---
viewed as spins, this is to say that if $S_z\approx +1/2$, then
$S_x$ will be small.
Hence, the system becomes uncondensed near total inversion.
If the excitation density is greater than the density of two-level
systems, the mean-field theory requires a coherent photon density.
Hence, the system condenses again, giving a re-entrant boundary.
Without saturable excitons, neither the uncondensing due to reduction
of mean-field polarisation, or the re-condensation due to exciton
saturation will occur.
With fluctuations, the multi-valued phase boundary is analogous to the
mean-field re-entrance, and appears to require saturable excitations
in the same manner.
Therefore, we do not expect such multi-valued phase boundaries in a
general model with a continuum of exciton states.

Let us now consider the case where there is a finite density of
exciton states, separated by the exciton binding energy from a
continuum of electron-hole states.
In such a case, if the continuum is well separated from bound states,
it may only affect the phase boundary at densities larger than those
where the features discussed above occur.
Well separated here means that the exciton binding energy is much
larger than the energy scales in our model, in particular, much larger
than $g\sqrt{n}$.
If the continuum only has effects at very high densities, the exotic
multi-valued phase boundary described in previous sections will be
realisable.
For the systems studied by Yamamoto {\it et al.}\cite{deng03:_polar},
$g\sqrt{n}\approx 7 \mathrm{meV}$, and the exciton binding energy
$Ryd^{\ast}\approx 10 \mathrm{meV}$.
For Dang {\it et al.}\cite{dang98:_stimul}, $g\sqrt{n}\approx
13\mathrm{meV}$, and $Ryd^{\ast} \approx 25 \mathrm{meV}$.
In neither case can the effects of the continuum be avoided.
Reducing the Rabi splitting, $g\sqrt{n}$, might allow the multi-valued
phase boundary to be observed.
However, reducing the Rabi splitting will decrease the transition
temperature in the region of interest.
In addition, to have well strong coupling, the Rabi splitting must
remain larger than the polariton linewidth due to photon lifetimes.

The existence of a photon dominated region at high densities is
however generic, and does not rely on a model with a finite total
density of states.
In an electron-hole plasma
model\cite{marchetti04:_conden_cavit_polar_disor_envir,marchetti04:comparison},
such a regime is also predicted.
At large densities, as the chemical potential (lying within the
exciton band) approaches the bottom of the photon band, the photon
density increases much faster than the exciton density.
The result in ref.~\onlinecite{marchetti04:_conden_cavit_polar_disor_envir}
only describes a photon dominated regime at zero temperature.
For the two-level system model, at high densities, the system remains
photon dominated, and with increasing temperature changes from coherent
to incoherent photons.
Such a change from coherent to incoherent photons is also expected to
occur in the electron-hole model, due to the large occupation of
bosonic modes near the chemical potential, but further work is
required here.

This discussion of how our model relates to experimental systems has
so far concentrated on what happens at high densities.
At lower densities (including the densities of current experiments),
no such significant differences are expected.
This is because, at low densities, higher energy exciton modes would
not be occupied, even if they exist.
In particular, the angular distribution of radiation
(figure~\ref{fig:nofk}) and excitation spectrum
(figures~\ref{fig:simple_spectrum} and~\ref{fig:dos}) should remain
unchanged for generic models.
Such signatures should therefore be expected for equilibrium
condensation in the systems currently studied.

The signatures of condensation presented in this paper are calculated
for a system in thermal equilibrium, while current experiments are
pumped.
For non-resonantly pumped
experiments\cite{weihs04:_polar,deng03:_polar,yamamoto02:_condensation,richard04:_angle_cdte_ii_vi,dang98:_stimul},
one must consider how pumping will modify the excitation spectrum and
the occupation of modes.
This can be described by coupling the system to baths describing
pumping of excitons and decay of photons.
For systems near equilibrium, with small coupling to baths, one expects the
excitation spectrum to remain close to the equilibrium case, but with
non-thermal occupations.
Even with non-thermal occupation, the large density of states for
excitations near the chemical potential can be expected to produce a
peak in the angular distribution of radiation.
For strong coupling to baths, the spectrum of excitations will also
change.
One particularly significant change is the possibility of giving a finite
lifetime to the Goldstone mode.
In this strongly pumped region, the signatures predicted in this paper
can be expected to change significantly.

It is also of interest to discuss how these signatures are related to
the behaviour seen in resonantly pumped
cavities\cite{baumberg02:_polar,savvidis00:_angle,baumberg00:_param}.
In these experiments pumping at a critical angle excites polaritons at
momentum $k_p$, causing emission from signal, $k=0$, and idler,
$k=2k_p$, modes.
Such a system may be described as an optical parametric oscillator.
Above a threshold, the luminescence from the signal increases
superlinearly, and the signal linewidth narrows dramatically.
In such a system, the occupation of the signal mode obeys a
self-consistency condition, and the relative phase between the pump
and signal mode is free.
However, the nature of this self-consistency differs from that for an
equilibrium condensate, and thus the signatures of equilibrium
condensation are no longer immediately applicable.

The form of the condensed luminescence spectrum, and the angular
distribution of polaritons depend on the existence of the low energy
Goldstone mode.
The energy of this mode vanishes as $k\rightarrow 0$ as a consequence
of the gap equation, eq.~(\ref{eq:7}), which means that global phase
rotations cost no energy.
In a laser, the coherent field is also set by a self-consistency
condition, balancing pumping and decay.
Like condensation, the laser transition can also be described as
symmetry breaking\cite{haken75}.
However, because the self-consistency relates imaginary parts of the
self energy, the dynamics of modes near the lasing mode is diffusive.
Therefore a free global phase and a self-consistency condition are not
sufficient for the signatures described in this paper.

For the optical parametric oscillator experiments, the
self-consistency equation is complicated by the existence of a
coherent idler field\cite{whittaker01:_class,eastham03:_stead}.
Since such experiments are strongly pumped it is expected that,
despite the free phase and self-consistency, the luminescence spectrum
and angular distribution of radiation as described in this paper will
not be applicable.
The laser and the equilibrium condensate are extreme cases, and the
distinction in practice is less
clear\cite{eastham03:_phase,baumberg00:_param}.
For example, adding decoherence\cite{szymanska02,szymanska03:_polar}
to an equilibrium condensate causes a crossover to a regime
better described as a laser.

\section{Conclusions}
\label{sec:conclusions}

We have studied the effect of fluctuations about the mean-field theory
for a model of localised excitons coupled to a continuum of photon
modes.
When condensed, the presence of a gap in the fermion density of
states, and the existence of the phase mode, cause dramatic changes
to the spectrum of collective modes.
Such changes lead to signatures of condensation in the luminescence
and absorption spectrum (figure~\ref{fig:dos}), and in the momentum
distribution of radiation escaping the cavity (figure~\ref{fig:nofk}).

Including the contribution to density due to fluctuations, we have
studied the crossover from a BEC of polaritons at low densities,
through a BCS-like mean-field regime at intermediate densities, and
finally to a BEC of massive photons at high densities, as shown in
figure~\ref{fig:phasediag}.
The BCS-like regime occurs at densities achieved in current
experiments.
Our study of the crossover can be compared to other studies of the
crossover; in Feshbach resonance
systems~\cite{ohashi02:_bcs_bec,ohashi03:_super_fermi_feshb}, or
for fermions interacting via a static four-particle
interaction\cite{nozieres85:_crossover,randeria:_cross}.
In distinction to those systems, due to the nature of the fermion
density of states in our system, there is no clear difference between
the r\^{o}les of the number and gap equations in the BEC and 
BCS-like regimes.  
Rather, the crossover is in the nature of the fluctuations that
depopulate the condensate.

At low densities, fluctuation corrections significantly alter the form
of the phase boundary from its mean-field form, leading to a
dependence $T\propto \rho$.
As the density increases, the transition temperature approaches the
Rabi splitting, and those single particle excitations which are
included in the mean-field calculation dominate, leading to a recovery
of the mean-field limit.
This occurs at a density scale set by the wavelength of light, not the
exciton separation, and so at densities much less than the Mott
density.
At yet higher densities, the system becomes photon dominated and the
boundary is that for a BEC of massive photons.
If the exciton is detuned below the photon, the mean-field boundary
can become multi-valued.
With fluctuations, this multi-valued structure occurs for smaller
detunings than are required for it to occur in the mean-field theory.
At least for a saturable two-level systems, this multi-valued boundary
likely indicates a first-order transition between two condensed
states.

Because our system is two-dimensional, it required fluctuations to be
considered in the presence of the condensate.
In the presence of a condensate it is important to consider changes to
the density both due to ``condensate depletion'' as well as the
occupation of fluctuation modes.
Such condensate depletion is included by taking full derivatives of
the action w.r.t.\ the chemical potential.

\section{Acknowledgements}
\label{sec:acknowledgements}

We are grateful to H.~Deng, L.~S.~Dang, F.~M.~Marchetti, B.~D.~Simons
and Y.~Yamamoto for useful discussions.  We acknowledge financial
support from the Cambridge-MIT Institute (JK), Sidney Sussex college,
Cambridge (PRE), Gonville and Caius College Cambridge (MHS) and the EU
Network ``Photon mediated phenomena in semiconductor nanostructures''
HPRN-CT-2002-00298.

\appendix

\section{Lehmann representation and broken symmetry}
\label{sec:lehm-repr-delt}

\subsection{Analytic properties of thermal and retarded greens functions}
\label{sec:analyt-prop-therm}

The inverse thermal Green's function contains terms proportional to
$\delta_{\omega}$.  Working from the Lehmann representation, it may be
shown that such terms can arise in the thermodynamic Green's function,
but not in the dynamic Green's functions.

\begin{widetext}
  To see this, consider the standard Lehmann representation
  (see {\it e.g.}\ ref.~\onlinecite{AGD}, section 17) 
  for the retarded greens function for Bose fields:
  \begin{eqnarray}
    \label{eq:81}
    G_{\mathrm{R}}(\omega)
    &=&
    \lim_{\delta\rightarrow 0^+}
    \int_{-\infty}^{\infty} \frac{\rhol(x) dx}{x-(\omega+i\delta)},
    \qquad 
    \rhol(x)=    \left( 1 - e^{-\beta x} \right) \sum_{n,m} 
    \left|\langle n \left| \psi \right| m \rangle \right|^2
    e^{\beta(F-E_n)}  \delta(x-E_{m n}).
  \end{eqnarray}
  However, for the thermal Green's function, an extra term may appear,
  \begin{eqnarray}
    \label{eq:83}
    \mathcal{G}(\omega)
    &=&
    \int_0^{\beta} d\tau e^{i\omega \tau} 
    \mathrm{Tr}\left(
      e^{\beta (F-H)}
      e^{H \tau} \psi e^{- H \tau} \psi^{\dagger}
    \right)
    =
    \sum_{n,m} 
    \left|\langle n \left| \psi \right| m \rangle \right|^2
    e^{\beta(F-E_n)} 
    \int_0^\beta e^{(i\omega - E_{m n}) \tau} d \tau
    \\\label{eq:84}
    &=&
    \int_{-\infty}^{\infty} \frac{\rhol(x) dx}{x+i\omega}
    +
    \beta \delta_\omega \sum_{n,m}
    \left|\langle n \left| \psi \right| m \rangle \right|^2
    e^{\beta(F-E_n)} 
    \delta(E_{m n}).
  \end{eqnarray}
\end{widetext}

The last term in (\ref{eq:84}) can be identified as the contribution
to the Green's function due to transitions between degenerate states;
or due to a macroscopic occupation of the photon in the ground state.
Such a term does not occur for the retarded Green's function, and so
in calculating the spectral Lehmann density $\rho_L(x)$, one may
neglect its effects.

\subsection{Matsubara summation with thermal greens functions}
\label{sec:mats-summ-with}

The $\delta_{\omega}$ terms will contribute to Matsubara
sums involving the thermal greens functions, in properties
such as the density.
In performing the Matsubara summation one must use
\begin{eqnarray}
  \label{eq:51}
  \sum_{\omega_n} f(\omega_n) &=& 
  \int_{-\infty}^{\infty} \frac{dz}{2\pi i} A(z) + f(0),
  \\
  A(z)&=& \lim_{\delta\rightarrow 0} 
  2 \Im \left[
    \tilde{f}(z^{\prime}) 
    \frac{\beta}{2}
    \coth \left( \frac{\beta z^{\prime}}{2} \right)
  \right]_{z^{\prime}=z+i\delta},
  \nonumber
\end{eqnarray}
where although $A(z)$ involves $\tilde{f}$, the analytic continuation
of $f$, $f(0)$ does not involve analytic continuation.
If the analytic continuation of $f$ is regular at $z=0$,  then
\begin{eqnarray}
  \label{eq:89}
  \sum_{\omega_n} f(\omega_n) 
  &=&
  \sum_{\mbox{poles of $f$}}
  \mathrm{res}
  \left[
    \tilde{f}(z) 
    \frac{\beta}{2}
    \coth \left( \frac{\beta z}{2} \right)
  \right]
  \nonumber\\
  &+&
  f(0) - \tilde{f}(0),
\end{eqnarray}
i.e.\ one must add a term to correct for the difference between $f$
and its analytic continuation at $z=0$.

\subsection{Emission and absorption coefficients}
\label{sec:emiss-absorpt-coeff}

Considering the Green's function for photon fluctuations; at zero
temperature the function $\rho_L(x)$ would give the density of states,
weighted by the photon component of a state.  At finite temperatures,
one may extract the probability to emit a photon,
\begin{eqnarray}
  \label{eq:85}
  P_{\mathrm{emit}}(x) &=& 
  \sum_{n,m} 
  \left|\langle m \left| \psi \right| n \rangle \right|^2
  e^{\beta(F-E_n)} \delta(x+E_{m n})
  \nonumber\\
  &=& n_{\mathrm{B}}(x) \rhol(x),
\end{eqnarray}
or to absorb a photon
\begin{eqnarray}
  \label{eq:86}
  P_{\mathrm{absorb}}(x) &=& 
  \sum_{n,m} 
  \left|\langle m \left| \psi^{\dagger} \right| n \rangle \right|^2
  e^{\beta(F-E_n)} \delta(x-E_{m n})
  \nonumber\\
  &=& (1+n_{\mathrm{B}}(x)) \rhol(x),
\end{eqnarray}

The energy $x$ is measured w.r.t.\ the chemical potential, so at zero
temperature there is only emission of photons at energies below the
chemical potential, or absorption of photons above the chemical
potential.
The Lehmann density itself, $\rhol(x)$ is the difference of these, and
can be interpreted as an absorption coefficient, which when negative
represents gain.

\end{document}